\documentclass{article} % For LaTeX2e

\usepackage{iclr2027_conference,times}

\usepackage{amsmath,amsfonts,bm}

\def\eqref#1{equation~\ref{#1}}
\def\1{\bm{1}}

\DeclareMathAlphabet{\mathsfit}{\encodingdefault}{\sfdefault}{m}{sl}
\SetMathAlphabet{\mathsfit}{bold}{\encodingdefault}{\sfdefault}{bx}{n}

\usepackage{listings}
\usepackage{textcomp}
\usepackage{xcolor}
\usepackage[scaled=0.8]{beramono}

\lstdefinestyle{numbers}{
    numbers=left,
    framexleftmargin=20pt,
    numberstyle=\tiny,
    firstnumber=auto,
    numbersep=1em,
    xleftmargin=2em
}

\lstdefinestyle{layout}{
    frame=none,
    captionpos=b,
}

\lstdefinestyle{default-style}{
    basicstyle=\fontencoding{T1}\ttfamily\footnotesize,
    style=numbers,
    style=layout,
    tabsize=2,
    upquote=true
}

\lstdefinelanguage{BASIC}{
    language=C++,
    style=default-style
}[keywords,comments,strings]

\usepackage{hyperref}
\usepackage{url}
\usepackage{xspace}
\usepackage{booktabs}
\usepackage{graphicx}
\usepackage[T1]{fontenc}
\usepackage{listings}
\usepackage{colortbl}
\usepackage{cleveref}
\usepackage{multirow}
\usepackage{capt-of}
\usepackage{wrapfig}
\usepackage{xcolor}
\usepackage{caption}
\usepackage{float}

\definecolor{diffincl}{rgb}{0.0, 0.5, 0.0}
\definecolor{diffrem}{rgb}{0.75, 0.0, 0.0}
\definecolor{diffstart}{rgb}{0.45, 0.45, 0.45}

\lstdefinelanguage{diff}{
    morecomment=[f][\color{diffrem}]{-},
    morecomment=[f][\color{diffincl}]{+},
    morecomment=[f][\color{diffstart}]{@}
}

\title{\multiswt: A Multilingual Benchmark for Reproduction Test Generation}

\author{
\textbf{Kazuki Kusama}\textsuperscript{1}\quad
\textbf{Sota Nakashima}\textsuperscript{1}\quad
\textbf{Haruka Tokumasu}\textsuperscript{1}\\
\textbf{Masanari Kondo}\textsuperscript{1}\quad
\textbf{Lingming Zhang}\textsuperscript{2}\quad
\textbf{Yasutaka Kamei}\textsuperscript{1}\\
\textsuperscript{1}Kyushu University\qquad
\textsuperscript{2}University of Illinois Urbana-Champaign\\
\texttt{\{kusama,nakashima,tokumasu\}@posl.ait.kyushu-u.ac.jp}\\
\texttt{\{kondo,kamei\}@ait.kyushu-u.ac.jp}\\
\texttt{lingming@illinois.edu}
}

\newcommand{\swt}{\textsc{SWT-Bench}\xspace}
\newcommand{\swe}{\textsc{SWE-Bench}\xspace}

\newcommand{\multiswe}{\textsc{Multi-SWE-Bench}\xspace}
\newcommand{\multiswt}{\textsc{Multi-SWT-Bench}\xspace}
\newcommand{\swtlite}{\textsc{Multi-SWT-Bench-Lite}\xspace}

\definecolor{mygray}{gray}{0.8}

\iclrfinalcopy % Show the author names in the arXiv version.
\begin{document}

\maketitle
\lhead{} % Omit the conference-publication header in the arXiv version.

\begin{abstract}
Reproduction test generation translates a natural-language issue description into executable tests that fail on the original code and pass after the issue is resolved, providing executable evidence for verifying candidate patches.
Existing benchmarks are constructed for individual programming languages, preventing a unified evaluation across diverse programming ecosystems.
To address this limitation, we introduce \multiswt, a multilingual benchmark for reproduction test generation consisting of 1,963 instances across eight programming languages: Python, Java, TypeScript, JavaScript, Go, Rust, C, and C++.
Using this benchmark, we conduct an empirical study of state-of-the-art LLMs with four representative methods (MSWE-agent, MOpenHands, Codex, and Claude Code) and perform a failure analysis across programming languages.
Our evaluation reveals a systematic language gap.
Across every evaluated method and LLM, the success rate on Python exceeds the aggregate success rate across all languages, while C++ exhibits particularly low success rates.
Our failure analysis identifies both language-specific challenges arising from repository testing conventions and cross-language challenges in inferring implicit setup requirements and preserving the target behavior through iterative revisions.
These findings demonstrate the importance of multilingual evaluation and provide actionable directions for developing reproduction test generation methods that generalize across software ecosystems and reliably capture issue-specific behavior.
\end{abstract}

% Main text (9 pages for initial submission)
\section{Introduction}
\label{sec:introduction}

%% Motivation for reproduction test generation
% Software maintenance requires developers to continuously understand existing systems and diagnose and resolve issues that arise during software evolution~\citep{bettenburg2008fse,zimmermann2010tse}.
% Advances in Large Language Models (LLMs) have shown substantial promise in automating issue resolution by inspecting repositories, modifying code, and proposing candidate fixes~\citep{mundler2024nips,zan2026nips,zhang2024issta,wang2023emnlp}.
% However, existing test suites may not exercise the specific behavior described in an issue, meaning that a patch can satisfy all existing tests without correctly resolving the reported problem~\citep{ye2021emse,yang2017fse}.
% Therefore, validating issue resolution requires tests that exercise the reported behavior.
Large Language Model (LLM)-powered agents are increasingly capable of automating software engineering tasks, including repository-level issue resolution~\citep{mundler2024nips,zan2026nips,zhang2024issta}.
As these agents become more capable of generating candidate patches, reliably verifying their outputs is emerging as a central challenge in AI-assisted software engineering~\citep{lu2026ieeesoftware}.
Reliable issue resolution requires not only generating a plausible patch but also determining whether it actually addresses the behavior described in the issue~\citep{li2026arxiv}.
Testing provides a natural mechanism for such verification, yet existing test suites do not necessarily exercise the behavior targeted by a newly reported issue.
Consequently, a candidate patch may pass all existing tests without actually resolving the reported problem~\citep{ye2021emse,yang2017fse}.

%% Introduction to reproduction test generation
Reproduction tests address this verification gap by translating the behavior described in a natural-language issue report into executable checks that fail on the original code and pass once the issue is resolved~\citep{mundler2024nips,ahmed2026arxiv}.
By reliably reproducing the reported failure, they provide execution-based feedback for investigating the issue and diagnosing its cause~\citep{nashid2025arxiv}.
Once a candidate patch is generated, these tests help determine whether the patch resolves the reported issue and, when retained in the test suite, detect future regressions that reintroduce the same problem~\citep{mundler2024nips}.

%% Motivation for multilingual evaluation
Existing benchmarks of reproduction test generation primarily focus on individual programming languages, such as Python~\citep{mundler2024nips,ahmed2024arxiv} and Java~\citep{ahmed2026arxiv}, leaving unclear whether conclusions about the capabilities and relative effectiveness of different agents generalize across programming ecosystems.
These ecosystems shape how tests are written, integrated, and executed, potentially exposing agents to different failure modes.
Systematic multilingual evaluation is therefore needed to assess the generalizability of existing agents, distinguish shared bottlenecks from language-specific challenges, and identify transferable design principles.

%% Introduction for our benchmark
To enable such an evaluation, we introduce \multiswt, a multilingual benchmark for reproduction test generation consisting of 1,963 instances across eight programming languages: Python, Java, TypeScript, JavaScript, Go, Rust, C, and C++.
% Following SWT-bench~\citep{mundler2024nips}, each task requires generating tests from an issue description and the corresponding pre-fix repository that fail before the issue is fixed and pass after it is resolved.
% We construct \multiswt from real-world issue-resolving instances and retain only those whose developer-written tests introduced with the original fix distinguish the pre-fix and post-fix versions.
Using \multiswt, we evaluate state-of-the-art models with four representative methods (MSWE-agent\footnote{\url{https://github.com/multi-swe-bench/MSWE-agent}}~\citep{zan2026nips}, MOpenHands\footnote{\url{https://github.com/multi-swe-bench/MopenHands}}~\citep{zan2026nips}, Codex\footnote{\url{https://openai.com/codex/}}, and Claude Code\footnote{\url{https://docs.anthropic.com/en/docs/claude-code}}) and conduct a detailed failure analysis.
Our empirical study reveals a systematic language gap: Python is the only language whose success rate consistently exceeds the aggregate success rate across all evaluated configurations, while C++ exhibits particularly low success rates.
% It also shows that triggering a failure or reaching code modified by the golden patch does not guarantee correct reproduction of the reported behavior.
Our failure analysis identifies both language-specific challenges rooted in ecosystem and repository conventions and cross-language challenges involving implicit setup requirements and preserving the target behavior through iterative revisions.

In summary, our main contributions are:

\begin{itemize}
    \item \textbf{\multiswt}: We introduce \multiswt, a multilingual benchmark for repository-level reproduction test generation, consisting of 1,963 instances across eight programming languages. It enables the systematic comparison of reproduction test generation performance across programming ecosystems under a unified evaluation setting.
    \item \textbf{Large-scale empirical evaluation}: We evaluate state-of-the-art LLMs across four representative methods on \multiswt, revealing their capabilities and limitations in reproduction test generation across programming languages.
    \item \textbf{Failure analysis}: We conduct a detailed failure analysis to identify language-specific and cross-language challenges, providing insights to guide future improvements in reproduction test generation methods for diverse software ecosystems.
\end{itemize}

\section{\multiswt}
\label{sec:multiswt}

\multiswt is a multilingual benchmark for reproduction test generation spanning eight programming languages: Python, Java, TypeScript, JavaScript, Go, Rust, C, and C++.
% Although derived from \multiswe~\citep{zan2026nips}, \multiswt evaluates a distinct capability: \multiswe evaluates agent-generated fixes against developer-written tests, whereas \multiswt requires agent-generated tests to fail before and pass after the developer-written fix.
% Because \multiswe neither requires nor separately evaluates generated tests, its resolution rate cannot isolate reproduction test generation capability.
% Following \swt~\citep{mundler2024nips}, we first introduce the notation used throughout the benchmark.
We then describe the construction process of \multiswt and analyze its key features.

\subsection{Notation and Definitions}
\label{sec:notation}

We denote a codebase \(\mathcal{R}\) after applying a patch \(X\) as \(\mathcal{R} \circ X\).
Given a single test \(t\) and a codebase \(\mathcal{R}\), executing \(t\) on \(\mathcal{R}\) results in either pass or fail.
We define this execution as \(\mathrm{exec}(t,\mathcal{R}) \in \{P,F\}\), where \(P\) and \(F\) denote passing and failing outcomes, respectively.
A generated test \(t\) successfully reproduces issue \(I\) on codebase \(\mathcal{R}\) if it fails on the original code (i.e., \(\mathrm{exec}(t,\mathcal{R})=F\)) but passes on the patched codebase (i.e., \(\mathrm{exec}(t,\mathcal{R}\circ X)=P\)).
We refer to such a test as a \emph{fail-to-pass} (\(F \rightarrow P\)) test.
For a generated test set \(\mathcal{T}\), we consider the issue successfully reproduced if \(\mathcal{T}\) contains at least one \(F \rightarrow P\) test and all generated tests pass on the patched codebase.

\subsection{Benchmark Construction}
\label{sec:benchmark-construction}

\paragraph{Data Source.}
We use 2,132 issue-resolution instances from~\multiswe~\citep{zan2026nips} to construct~\multiswt.
These instances are derived from real-world GitHub issues and their corresponding issue-resolving pull requests.
Each source instance provides the issue description, the pre-fix codebase $\mathcal{R}$, the issue-resolving code changes (\emph{golden patch} $X^*$), and the developer-written tests introduced with the fix (\emph{golden reference tests} $T^*$).

\newpage

\paragraph{Construction of~\multiswt.}
Starting from these issue-resolution instances, we construct~\multiswt in two steps: task conversion and instance validation.

\textit{(1) Task conversion.}
For each source instance, we retain the issue description and the pre-fix codebase $\mathcal{R}$ as inputs and change the generation target from a code patch to a test set $\mathcal{T}$.
Generated tests are executed on both $\mathcal{R}$ and the codebase after applying the golden patch $\mathcal{R}\circ X^*$.
A generated test reproduces the issue when it fails on the pre-fix codebase and passes after the fix, following the $F \rightarrow P$ criterion defined in Section~\ref{sec:notation}.

\textit{(2) Instance validation.}
We validate whether the developer-written golden reference tests $T^*$ provide a reliable oracle for the reproduction test generation task.
We execute $T^*$ on both $\mathcal{R}$ and $\mathcal{R}\circ X^*$ and retain only instances that contain at least one $F \rightarrow P$ reference test and for which all reference tests pass after applying $X^*$.
Instances whose reference-test outcomes cannot be reliably evaluated are also excluded.
This process removes 169 of the 2,132 source instances, resulting in~\multiswt with 1,963 instances.
We further construct~\swtlite, a 340-instance subset of~\multiswt for lower-cost evaluation.
We use~\swtlite for all experiments reported in this paper.

\subsection{Features of~\multiswt}
\label{sec:benchmark-features}
\begin{wraptable}{R}{0.5\columnwidth}
    \centering
    \caption{Composition of~\multiswt.}
    \label{tab:multiswt-composition}
    \small
    \setlength{\tabcolsep}{4pt}
    \begin{tabular}{lcc}
        \toprule
        Language & \# Repositories & \# Instances \\
        \midrule
        Python     & 12 & 474 \\
        Java       & 9  & 111 \\
        TypeScript & 3  & 224 \\
        JavaScript & 6  & 356 \\
        Go         & 3  & 425 \\
        Rust       & 10 & 120 \\
        C          & 3  & 127 \\
        C++        & 5  & 126 \\
        \midrule
        Total      & 51 & 1,963 \\
        \bottomrule
    \end{tabular}
\end{wraptable}

\multiswt contains 1,963 reproduction test generation instances from 51 repositories across eight programming languages, with at least 111 instances per language (Table~\ref{tab:multiswt-composition}).
As shown in Table~\ref{tab:multiswt-characteristics}, the benchmark spans substantial variation in issue context, repository scale, and reference-test characteristics; for example, the average repository size ranges from 93K to 994K lines of code across languages.
The golden reference tests contain 17.3 test cases per instance on average, of which 15.1 (approximately 87\%) exhibit the $F \rightarrow P$ behavior, indicating that most directly distinguish the buggy version from the issue-resolving version.
These characteristics provide a diverse multilingual evaluation set spanning different software ecosystems and task contexts.

\begin{table*}[h]
    \centering
    \caption{Characteristics of~\multiswt. Values are averages within each language. Golden reference tests are counted at the executed test-case level.}
    \label{tab:multiswt-characteristics}
    \small
    \setlength{\tabcolsep}{3pt}
    \begin{tabular}{lcccccccc}
            \toprule
            & \multicolumn{1}{c}{Issue Description}
            & \multicolumn{2}{c}{Codebase}
            & \multicolumn{3}{c}{Golden Reference Tests}
            & \multicolumn{2}{c}{Test Patch} \\
            \cmidrule(lr){2-2}
            \cmidrule(lr){3-4}
            \cmidrule(lr){5-7}
            \cmidrule(lr){8-9}
            Language & \# Words & \# Files & \# LoC (K) & \# Tests & \# F$\rightarrow$P & \# P$\rightarrow$P & \shortstack{\# Files \\ Changed} & \shortstack{\# Lines \\ Changed} \\
            \midrule
            Python     & 184.8 & 3,185.9  & 704.1 & 3.9  & 2.2  & 1.7 & 1.3 & 23.9 \\
            Java       & 234.0 & 3,269.0  & 903.6 & 24.2 & 22.5 & 1.7 & 1.5 & 79.4 \\
            TypeScript & 199.5 & 26,561.2 & 827.7 & 16.4 & 16.3 & 0.0 & 2.0 & 91.2 \\
            JavaScript & 137.7 & 5,611.1  & 167.8 & 4.6  & 4.6  & 0.0 & 5.1 & 86.8 \\
            Go         & 152.0 & 546.6    & 101.6 & 20.1 & 12.9 & 7.2 & 2.4 & 123.0 \\
            Rust       & 245.0 & 504.5    & 93.2  & 76.1 & 75.8 & 0.3 & 3.5 & 122.3 \\
            C          & 163.1 & 623.7    & 161.3 & 33.3 & 31.1 & 2.2 & 1.9 & 112.5 \\
            C++        & 212.7 & 734.9    & 994.3 & 18.2 & 18.2 & 0.0 & 2.8 & 294.3 \\
            \midrule
            Overall    & 177.7 & 5,239.4  & 448.0 & 17.3 & 15.1 & 2.2 & 2.6 & 96.7 \\
            \bottomrule
    \end{tabular}
\end{table*}

\section{Experimental Setup}
\label{sec:experimental-setup}

\subsection{Evaluated Methods and LLMs}
\label{sec:evaluated-methods-llms}

\textbf{Methods.}
We evaluate four representative methods for reproduction test generation: MSWE-agent\footnote{\url{https://github.com/multi-swe-bench/MSWE-agent}}, MOpenHands\footnote{\url{https://github.com/multi-swe-bench/MopenHands}}, Codex\footnote{\url{https://openai.com/codex/}}, and Claude Code\footnote{\url{https://docs.anthropic.com/en/docs/claude-code}}.
MSWE-agent and MOpenHands are issue-resolution agents used in \multiswe~\citep{zan2026nips}, whereas Codex and Claude Code are general-purpose coding agents.
We adapt all four methods to reproduction test generation.

\begin{itemize}
    \item \textbf{MSWE-agent}: MSWE-agent is built to support multilingual adoption based on SWE-agent\footnote{\url{https://github.com/SWE-agent/SWE-agent}}~\citep{yang2024nips}, which is an agent-based approach that solves issues through multi-turn interactions via a predefined agent-computer interface.
    We replace its issue-resolution prompts with prompts for reproduction test generation and extend its execution harness to use the repository-specific configurations provided by \multiswt.
    
    \item \textbf{MOpenHands}: MOpenHands is based on OpenHands\footnote{\url{https://github.com/All-Hands-AI/OpenHands}}, which is a widely adopted platform for building software development agents.
    Following SWT-bench~\citep{mundler2024nips}, we adapt its prompt for reproduction test generation and disable browsing and external network access except for connections to the LLM service.

    \item \textbf{Codex}: Codex is an AI coding agent developed by OpenAI.
    We use its command-line version with the issue description and a task prompt based on the MSWE-agent prompt.
    Each instance is processed in a fresh container and agent session, with web search disabled and external network access blocked except for communication with the model service.

    \item \textbf{Claude Code}: Claude Code is an agentic coding tool developed by Anthropic.
    We use its command-line version with the same task prompt and execution-isolation settings as Codex, and additionally disable web fetching.
\end{itemize}

The prompts used for all methods are provided in Appendix~\ref{app:prompts}.

\textbf{LLMs.}
% We evaluate MSWE-agent and MOpenHands with GPT-5.6 Luna, Claude Sonnet 5, DeepSeek-V4-Flash, and DeepSeek-V4-Pro; Codex with GPT-5.6 Luna and GPT-6 Astra; and Claude Code with Claude Sonnet 5 and Claude Opus 5.
For MSWE-agent and MOpenHands, which access LLMs through usage-based APIs, inference cost is an important practical constraint at the scale of our evaluation. 
We therefore select competitive models from multiple providers and price tiers to cover different cost--performance trade-offs.
Codex and Claude Code are evaluated under subscription plans. 
For these methods, we include models shared with MSWE-agent and MOpenHands to enable cross-method comparison, as well as the more capable models available through their respective plans (GPT-6 Astra and Claude Opus 5) to broaden the range of models evaluated within our experimental budget.

\subsection{Evaluation Metrics}
\label{sec:evaluation-metrics}
Following \swt~\citep{mundler2024nips}, we adopt three metrics to evaluate the performance of any method: success rate ($\mathcal{S}$), change coverage ($\Delta\mathcal{C}$), and patch well-formedness ($\mathcal{W}$).

\textbf{Success Rate.}
% We measure the success rate $\mathcal{S}$ as the proportion of instances in which the generated test set $\mathcal{T}$ successfully reproduces the reported issue.
% An instance is considered successful if $\mathcal{T}$ contains at least one Fail-to-Pass ($F \rightarrow P$) test and no test that fails after applying the golden patch ($\times \rightarrow F$).
We measure the success rate $\mathcal{S}$ as the proportion of instances in which the generated test set $\mathcal{T}$ successfully reproduces the reported issue, requiring at least one Fail-to-Pass ($F \rightarrow P$) test and no test that fails after applying the golden patch ($\times \rightarrow F$).
We additionally report the proportions of instances for which the generated test set contains at least one Fail-to-Pass ($F \rightarrow P$), Fail-to-Any ($F \rightarrow \times$), and Pass-to-Pass ($P \rightarrow P$) test.
Here, $F \rightarrow \times$ represents a test that fails on the original codebase regardless of its outcome after applying the golden patch, while $P \rightarrow P$ represents a test that passes both before and after the patch.

\textbf{Change Coverage.}
We measure change coverage $\Delta\mathcal{C}$ as the proportion of executable lines modified by the golden patch that are additionally covered by the generated tests.
We consider only executable lines in the golden patch, excluding non-executable changes such as documentation or configuration files.
A changed line is considered executable if it is executed by either the original test suite $\mathcal{T}^{\mathcal{R}}$ or the golden reference test suite $\mathcal{T}^{*}$.
We separately consider removed or modified lines in the original codebase $\mathcal{R}$ and added or modified lines in the patched codebase $\mathcal{R} \circ X^{*}$.
Formally, let
$\mathcal{C}^{\mathcal{R}}_{\mathcal{T}}(l) \in \mathbb{Z}^{\geq 0}$
denote the number of times line $l$ is executed when running test suite $\mathcal{T}$ on codebase $\mathcal{R}$.
We define the executable removed and added lines of the golden patch $X^{*}$ as
\begin{align}
    \mathcal{X}^{*}_{r}
    &=
    \{ l \in X^{*}_{r}
    \mid
    \mathcal{C}^{\mathcal{R}}_{\mathcal{T}^{\mathcal{R}}}(l)
    +
    \mathcal{C}^{\mathcal{R}}_{\mathcal{T}^{*}}(l)
    > 0
    \}, \\
    \mathcal{X}^{*}_{a}
    &=
    \{ l \in X^{*}_{a}
    \mid
    \mathcal{C}^{\mathcal{R} \circ X^{*}}_{\mathcal{T}^{\mathcal{R}}}(l)
    +
    \mathcal{C}^{\mathcal{R} \circ X^{*}}_{\mathcal{T}^{*}}(l)
    > 0
    \},
\end{align}
where $X^{*}_{r}$ and $X^{*}_{a}$ denote the lines removed and added by the golden patch, respectively.
The change coverage of the generated tests $\mathcal{T}$ is then defined as
\begin{equation}
    \Delta \mathcal{C}^{X^{*}}_{\mathcal{T}}
    =
    \frac{
        |\{ l \in \mathcal{X}^{*}_{r}
        \mid
        \mathcal{C}^{\mathcal{R}}_{\mathcal{T}^{\mathcal{R}} \cup \mathcal{T}}(l)
        >
        \mathcal{C}^{\mathcal{R}}_{\mathcal{T}^{\mathcal{R}}}(l)
        \}|
        +
        |\{ l \in \mathcal{X}^{*}_{a}
        \mid
        \mathcal{C}^{\mathcal{R} \circ X^{*}}_{\mathcal{T}^{\mathcal{R}} \cup \mathcal{T}}(l)
        >
        \mathcal{C}^{\mathcal{R} \circ X^{*}}_{\mathcal{T}^{\mathcal{R}}}(l)
        \}|
    }{
        |\mathcal{X}^{*}_{r}| + |\mathcal{X}^{*}_{a}|
    }.
\end{equation}

% We report the average change coverage over all applicable instances as $\Delta\mathcal{C}_{\mathrm{all}}$, and separately over successful and non-successful instances as $\Delta\mathcal{C}_{\mathcal{S}}$ and $\Delta\mathcal{C}_{\neg\mathcal{S}}$, respectively.

We exclude instances with no executable lines modified by the golden patch, i.e.,
$|\mathcal{X}^{*}_{r}| + |\mathcal{X}^{*}_{a}| = 0$.

\textbf{Patch Well-Formedness.}
We measure patch well-formedness $\mathcal{W}$ as the proportion of instances for which the generated patch can be successfully applied to the original codebase.
Formally, an instance is considered well-formed if the generated patch $\mathcal{X}$ can be applied to the original codebase $\mathcal{R}$ without errors.
Since a generated test can be executed only after the patch is successfully applied, patch well-formedness captures a necessary condition for evaluating test-generation performance.

\section{Experimental Results}
\label{sec:experimental-results}

\subsection{Performance on \multiswt}
\label{sec:performance-on-multiswt}
We evaluate reproduction test generation performance on \multiswt along two dimensions: (1) language-specific performance and (2) comparisons across LLMs and agents.

\subsubsection{Performance across Programming Languages}
\label{sec:performance-languages}
% Table~\ref{tab:overall-performance} presents the overall performance across eight programming languages.
% Based on the results, several key observations can be drawn and outlined below.

\begin{table*}[ht]
    \centering
    \renewcommand{\arraystretch}{1.1}
    \caption{Success rate (\%) of reproduction test generation on \multiswt.}
    \label{tab:overall-performance}
    \small
    \begin{tabular}{lccccccccc}
        \toprule
        \textbf{Models} & \textbf{All} & \textbf{Python} & \textbf{Java} & \textbf{TS} & \textbf{JS} & \textbf{Go} & \textbf{Rust} & \textbf{C} & \textbf{C++} \\
        \hline

        \rowcolor{mygray}\multicolumn{10}{c}{\textbf{MSWE-agent}} \\
        GPT-5.6 Luna      & 40.00 & 47.50 & 55.00 & 35.56 & 40.00 & 33.33 & 53.33 & 30.00 & 25.00 \\
        Claude Sonnet 5   & 31.18 & 42.50 & 42.50 & 22.22 & 37.78 & 17.78 & 44.44 & 25.00 & 17.50 \\
        DeepSeek-V4-Flash & 11.76 & 20.00 & 15.00 & 13.33 & 17.78 &  8.89 &  6.67 & 10.00 &  2.50 \\
        DeepSeek-V4-Pro   & 17.65 & 30.00 & 37.50 & 15.56 & 24.44 &  4.44 & 17.78 & 10.00 &  2.50 \\

        \midrule
        \rowcolor{mygray}\multicolumn{10}{c}{\textbf{MOpenHands}} \\
        GPT-5.6 Luna      & 47.35 & 57.50 & 50.00 & 66.67 & 44.44 & 33.33 & 44.44 & 47.50 & 35.00 \\
        Claude Sonnet 5   & 57.94 & 67.50 & 37.50 & 80.00 & 53.33 & 46.67 & 71.11 & 62.50 & 42.50 \\
        DeepSeek-V4-Flash & 55.29 & 62.50 & 45.00 & 73.33 & 53.33 & 46.67 & 66.67 & 50.00 & 42.50 \\
        DeepSeek-V4-Pro   & 48.24 & 50.00 & 42.50 & 64.44 & 46.67 & 42.22 & 53.33 & 47.50 & 37.50 \\

        \midrule
        \rowcolor{mygray}\multicolumn{10}{c}{\textbf{Codex}} \\
        GPT-5.6 Luna      & 43.53 & 62.50 & 40.00 & 55.56 & 42.22 & 31.11 & 42.22 & 42.50 & 32.50 \\
        GPT-6 Astra       & 56.47 & 67.50 & 50.00 & 73.33 & 57.78 & 44.44 & 57.78 & 55.00 & 45.00 \\

        \midrule
        \rowcolor{mygray}\multicolumn{10}{c}{\textbf{Claude Code}} \\
        Claude Sonnet 5   & 51.18 & 67.50 & 35.00 & 66.67 & 46.67 & 37.78 & 60.00 & 57.50 & 37.50 \\
        Claude Opus 5     & 65.29 & 82.50 & 50.00 & 66.67 & 66.67 & 62.22 & 71.11 & 70.00 & 52.50 \\

        \bottomrule
    \end{tabular}
\end{table*}

\textbf{Python is the only language whose success rate exceeds the aggregate success rate across all languages (All) for every evaluated method--LLM configuration.}
As shown in Table~\ref{tab:overall-performance}, Python exceeds All across all 12 configurations, whereas other languages have at least one configuration in which its success rate does not exceed All.
A similar advantage for Python has also been reported in issue resolution~\citep{zan2026nips} and multilingual code generation~\citep{ivanova2026iclr}.
Prior work has discussed the greater representation of Python in training data and Python-oriented method design as possible contributors to this trend.
Our results show that the same trend also emerges in reproduction test generation, suggesting that evaluations limited to Python may provide an overly optimistic view of reproduction test generation performance.

\textbf{C++ exhibits particularly low success rates across the evaluated method--LLM configurations.}
Table~\ref{tab:overall-performance} shows that Go and C++ are the only languages whose success rates remain below All across all 12 configurations.
Among these two languages, C++ achieves a lower success rate than Go in nine of the 12 configurations.
Even Claude Code + Claude Opus 5, which achieves the highest overall success rate, achieves only 52.50\% on C++.
Characteristics of the C++ instances may partly explain this lower success rate.
The average codebase size is 994.3K LoC, and the average number of changed lines in the test patches is 294.3, both the largest among the eight languages (Table~\ref{tab:multiswt-characteristics}).
These characteristics suggest that the low success rate on C++ may be associated not only with language-specific factors but also with properties of the C++ instances in our benchmark.

\subsubsection{Performance across Various Methods and LLMs}
\label{sec:performance-methods-llms}

\begin{table*}[ht]
    \centering
    \renewcommand{\arraystretch}{1.1}
    \caption{Overall reproduction test generation performance (\%) of different agents and LLMs on \multiswt.}
    \label{tab:method-llm-performance}
    \small
    \begin{tabular}{lcccccccc}
        \toprule
        \textbf{Models}
        & $\mathbf{W}$
        & $\mathbf{S}$
        & $\mathbf{F \rightarrow \times}$
        & $\mathbf{F \rightarrow P}$
        & $\mathbf{P \rightarrow P}$
        & $\mathbf{\Delta C_{\mathrm{all}}}$
        & $\mathbf{\Delta C_{S}}$
        & $\mathbf{\Delta C_{\neg S}}$ \\
        \hline

        \rowcolor{mygray}\multicolumn{9}{c}{\textbf{MSWE-agent}} \\
        GPT-5.6 Luna
        & 95.59 & 40.00 & 75.59 & 46.76 & 80.88 & 32.87 & 46.69 & 25.76 \\
        Claude Sonnet 5
        & 94.71 & 31.18 & 54.12 & 36.47 & 82.06 & 23.77 & 44.54 & 15.80 \\
        DeepSeek-V4-Flash
        & 93.82 & 11.76 & 25.59 & 13.82 & 85.59 & 12.44 & 41.75 & 9.27 \\
        DeepSeek-V4-Pro
        & 95.59 & 17.65 & 29.12 & 19.12 & 87.35 & 13.72 & 39.32 & 8.79 \\

        \midrule
        \rowcolor{mygray}\multicolumn{9}{c}{\textbf{MOpenHands}} \\
        GPT-5.6 Luna
        & 99.71 & 47.35 & 82.94 & 54.41 & 77.94 & 36.82 & 48.33 & 26.81 \\
        Claude Sonnet 5
        & 99.12 & 57.94 & 85.29 & 69.41 & 79.12 & 42.06 & 52.43 & 28.73 \\
        DeepSeek-V4-Flash
        & 98.53 & 55.29 & 83.82 & 64.71 & 78.82 & 41.50 & 53.10 & 28.53 \\
        DeepSeek-V4-Pro
        & 99.71 & 48.24 & 82.65 & 58.82 & 78.82 & 41.16 & 53.41 & 30.49 \\

        \midrule
        \rowcolor{mygray}\multicolumn{9}{c}{\textbf{Codex}} \\
        GPT-5.6 Luna
        & 100.00 & 43.53 & 81.76 & 53.24 & 77.94 & 37.12 & 45.53 & 31.52 \\
        GPT-6 Astra
        & 100.00 & 56.47 & 84.12 & 69.12 & 78.24 & 43.62 & 54.29 & 30.99 \\

        \midrule
        \rowcolor{mygray}\multicolumn{9}{c}{\textbf{Claude Code}} \\
        Claude Sonnet 5
        & 99.41 & 51.18 & 80.88 & 59.71 & 76.18 & 38.44 & 48.10 & 30.95 \\
        Claude Opus 5
        & 100.00 & 65.29 & 89.12 & 79.41 & 79.12 & 50.87 & 58.93 & 37.35 \\

        \bottomrule
    \end{tabular}
\end{table*}

\begin{figure}[t]
    \centering
    \includegraphics[width=\linewidth]{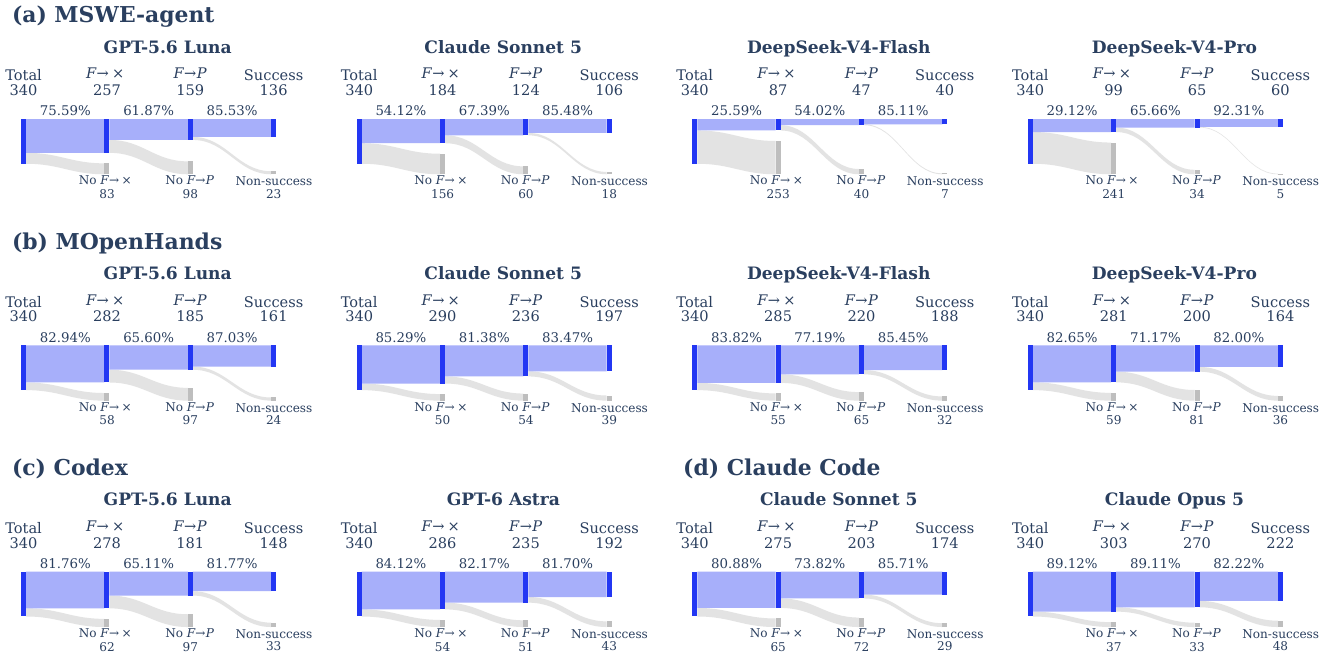}
    \caption{Progression of instances through F$\rightarrow\times$, F$\rightarrow$P, and success across methods and LLMs.}
    \label{fig:reproduction-test-generation-flow}
\end{figure}

% Table~\ref{tab:method-llm-performance} shows the overall performance for each evaluated method and LLM.

\textbf{For every LLM evaluated with multiple methods, MOpenHands achieves the highest success rate.}
Table~\ref{tab:method-llm-performance} compares four LLMs evaluated with multiple methods: GPT-5.6 Luna, Claude Sonnet 5, DeepSeek-V4-Flash, and DeepSeek-V4-Pro.
MOpenHands achieves the highest success rate for each of these LLMs.
% With Claude Sonnet 5, for example, MOpenHands achieves a success rate of 57.94\%, outperforming Claude Code at 51.18\% and MSWE-agent at 31.18\%.
A similar tendency for MOpenHands to achieve strong performance has also been reported for issue resolution~\citep{zan2026nips}, suggesting that its effectiveness extends to reproduction test generation.
One possible explanation is that MOpenHands and MSWE-agent interact with repositories differently.
MOpenHands can execute general-purpose commands and revise its actions based on execution feedback, whereas MSWE-agent uses a more structured interface~\citep{wang2025iclr,yang2024nips}.
This flexibility may help MOpenHands generate tests that correctly reproduce reported issues across different repositories.

\textbf{The best-performing LLM varies across reproduction test generation methods.}
As shown in Table~\ref{tab:method-llm-performance}, MSWE-agent and MOpenHands are evaluated with the same four LLMs but achieve their highest success rates with different LLMs.
GPT-5.6 Luna achieves the highest success rate with MSWE-agent (40.00\%), whereas Claude Sonnet 5 achieves the highest success rate with MOpenHands (57.94\%).
These results show that the LLM performing best with one method does not necessarily perform best with another, highlighting the importance of selecting an appropriate combination of method and underlying LLM for reproduction test generation.

\textbf{For MOpenHands, Codex, and Claude Code, a key challenge is generating tests whose failures are resolved by the golden patch.}
Figure~\ref{fig:reproduction-test-generation-flow} shows how instances progress through three stages: generating at least one test that fails on the buggy version (F$\rightarrow\times$), generating at least one test that fails on the buggy version and passes after applying the golden patch (F$\rightarrow$P), and ultimately satisfying the success criterion.
For MOpenHands, Codex, and Claude Code, all evaluated LLMs trigger at least one failure on the buggy version in more than 80\% of instances, but their F$\rightarrow$P rates are 9.71 to 28.53 percentage points lower than their corresponding F$\rightarrow\times$ rates.
In contrast, with MSWE-agent, the rate of triggering at least one failure ranges from 25.59\% to 75.59\% across LLMs, indicating that merely triggering a failure is itself a challenge for some LLMs under this method.

\textbf{Reaching code modified by the golden patch does not guarantee successful reproduction of the reported issue.}
As shown in Table~\ref{tab:method-llm-performance}, the average change coverage of non-successful instances ($\Delta C_{\neg S}$) is greater than 0\% for every evaluated method--LLM configuration.
In particular, for Claude Code + Claude Opus 5, $\Delta C_{\neg S}$ reaches 37.35\%.
These results show that unsuccessful generations can reach code modified by the golden patch, suggesting that correctly capturing the behavioral difference described in the issue after reaching such code remains a bottleneck.

We additionally analyze whether the performance is associated with (1) issue type, (2) issue description characteristics, (3) fix patch characteristics, and (4) reference test characteristics. 
% We additionally analyze whether the performance is associated with (1) issue type, (2) issue description length, (3) fix patch length, and (4) the number of files modified by the golden fix.
We report the detailed results in Appendix~\ref{app:influencing-factors}.
% \sota{TODO: need to add new factors.}

\subsection{Failure Analysis}
\label{sec:failure-analysis}

We examine instances in which the generated tests failed to reproduce the reported issue.
Through a detailed analysis of these instances, we identify challenges that are specific to particular programming languages, as well as challenges that occur across languages.
Detailed examples of the analyzed failures are provided in Appendix~\ref{app:failure-examples}.

\subsubsection{Language-Specific Challenges}
\label{sec:language-specific-challenges}

\textbf{Programming-language and ecosystem-specific testing conventions can cause reproduction test generation failures (Appendix~\ref{app:failure-testing-conventions}).}
In Rust, 18 of the 45 issues involve CLI output or diagnostics, making it important to align the command options used in a test with the output expected by that test.
For example, in \href{https://github.com/BurntSushi/ripgrep/issues/1642}{\texttt{ripgrep \#1642}}, the agent expected line-numbered output without enabling the command option that displays line numbers, so the test still failed after applying the golden patch.
% The same type of failure was also observed in \href{https://github.com/sharkdp/fd/issues/555}{\texttt{fd \#555}} and \href{https://github.com/sharkdp/bat/issues/1276}{\texttt{bat \#1276}}.
In Java, reusing existing helpers or fixtures can introduce conditions not required by the issue.
For example, in \href{https://github.com/FasterXML/jackson-databind/issues/1923}{\texttt{Jackson Databind \#1923}}, the reused helper checks for a specific exception message that is not required by the issue, causing the test to fail even after the reported problem is fixed.
% $\Rightarrow$ \textit{For developers, reproduction test generation agents should be designed to account for programming-language and ecosystem-specific testing conventions.}
% Agents should inspect how existing tests are used and how project-specific test setup is configured, and include only the conditions needed to reproduce the issue.

\textbf{A test that correctly reproduces an issue can still lead to test generation failure by affecting existing tests (Appendix~\ref{app:failure-existing-tests}).}
In C++, test-framework projects such as Catch2 can use repository-specific validation workflows in which newly added tests also affect existing tests.
For example, \href{https://github.com/catchorg/Catch2/issues/2719}{\texttt{Catch2 \#2719}} reports a case where an unexpected exception is not correctly reported as a test failure.
The agent generated four tests for this behavior, all of which passed after applying the golden patch.
However, Catch2 also has an existing test that runs many tests together and compares their combined output with saved expected output.
The newly added tests changed this output, causing the existing test to fail even after applying the golden patch and making the entire test set Non-success.

\subsubsection{Cross-Language Challenges}
\label{sec:cross-language-challenges}

\textbf{Even when an issue provides concrete inputs and reproduction steps, directly translating them into a test may not reproduce the bug (Appendix~\ref{app:failure-implicit-setup}).}
For example, \href{https://github.com/iamkun/dayjs/issues/1022}{\texttt{Day.js \#1022}} reports an incorrect relative-time result for a specific date and provides both the input date and the expected output.
However, reproducing the bug also requires controlling the current time and enabling the relevant plugins in the test environment.
% Similarly, \href{https://github.com/vuejs/core/issues/11624}{\texttt{Vue \#11624}} reports an error thrown in a computed property that is not handled by the component's error handler, and the issue provides a concrete action that triggers the error.
% Reproducing the same problem in a test, however, also requires setting up the component and watch so that the computed property is evaluated under the relevant conditions; the agent failed to produce a reproduction test with this setup.
This example highlights that reproduction test generation requires not only translating the reported behavior into assertions, but also identifying and instantiating the environmental conditions under which that behavior occurs.
% $\Rightarrow$ \textit{For developers, reproduction test generation agents should be designed to identify setup requirements beyond the inputs and steps explicitly stated in an issue.}
% Before generating a test, agents should inspect relevant code and existing tests to identify these requirements and incorporate them into the test setup.

\textbf{A reproduction test can lose its intended behavior or executability after subsequent revisions (Appendix~\ref{app:failure-revisions}).}
In \href{https://github.com/expressjs/express/issues/3695}{\texttt{Express \#3695}}, the agent first created a test that reproduced the reported failure on the buggy version successfully.
However, after revising the test to satisfy the project's lint rules, the test no longer exercised the behavior described in the issue and still failed after applying the golden patch.
Similarly, in \href{https://github.com/ponylang/ponyc/issues/1051}{\texttt{Ponyc \#1051}}, the agent initially created a test that failed on the buggy version, but a later edit introduced a duplicate test definition and prevented the final test suite from compiling.
These cases show that subsequent revisions can invalidate a previously working reproduction test and lead to test generation failure.
% $\Rightarrow$ \textit{For developers, reproduction test generation agents should be designed to revalidate tests after modifying them.}
% Before submission, agents should confirm that the revised test still reproduces the same issue and remains executable.

% \textbf{An agent can be classified as Non-success even when it generates tests consistent with the issue, if the golden patch addresses only part of the reported behavior.}
% For example, \href{https://github.com/clap-rs/clap/issues/5526}{\texttt{Clap \#5526}} reports two problems for an argument configured to require at least two values: (1) no error is raised when no value is provided, and (2) the error message is incorrect when one value is provided.
% However, the golden patch corrects the error message but does not address the zero-value case.
% The agent generated tests for both behaviors, so one passed after applying the golden patch while the other continued to fail, causing the test set to be classified as Non-success.
% In such cases, benchmark evaluation may underestimate agent performance because the golden patch covers only part of the behavior described in the issue.\\
% % $\Rightarrow$ \textit{For benchmark designers, evaluations should account for mismatches between the behaviors described in an issue and those addressed by the golden patch.}
% % They should distinguish tests that are unsupported by the issue from tests that check issue-described behavior not covered by the golden patch.

\subsection{Resource Consumption}
\label{sec:resource-consumption}

\begin{table*}[ht]
    \centering
    \caption{Average inference cost per instance (\$).}
    \label{tab:resource-consumption}
    \small
    \setlength{\tabcolsep}{3pt}
    \begin{tabular}{lcccc}
        \toprule
        Models
        & MSWE-agent
        & MOpenHands
        & Codex
        & Claude Code \\
        \midrule
        GPT-5.6 Luna
        & 0.0309 & 0.0294 & 0.0202 & -- \\
        GPT-6 Astra
        & -- & -- & 0.4429 & -- \\
        Claude Sonnet 5
        & 0.5634 & 0.6222 & -- & 0.5076 \\
        Claude Opus 5
        & -- & -- & -- & 1.1545 \\
        DeepSeek-V4-Flash
        & 0.0359 & 0.0175 & -- & -- \\
        DeepSeek-V4-Pro
        & 0.1370 & 0.0339 & -- & -- \\
        \bottomrule
    \end{tabular}
\end{table*}

Table~\ref{tab:resource-consumption} reports the average estimated inference cost per instance.
Dashes indicate configurations that were not evaluated in our experiments.
We calculate costs from recorded token usage using each model's official API pricing.
Codex and Claude Code were run under subscription plans, so we report their API-equivalent costs for comparison.

\textbf{With MOpenHands, DeepSeek-V4-Flash achieves a success rate close to Claude Sonnet 5 at a lower inference cost.}
DeepSeek-V4-Flash achieves a success rate of 55.29\% at \$0.0175/instance, compared with 57.94\% at \$0.6222/instance for Claude Sonnet 5.
Across all evaluated configurations, inference cost ranges from \$0.0175 to \$1.1545 per instance.
These results highlight the importance of considering inference cost together with success rate when selecting an LLM for reproduction test generation.

\textbf{For three of the four LLMs shared by MOpenHands and MSWE-agent, MOpenHands achieves both a higher success rate and a lower inference cost.}
With DeepSeek-V4-Pro, MOpenHands achieves a success rate of 48.24\% at \$0.0339/instance, compared with 17.65\% at \$0.1370/instance for MSWE-agent.
Claude Sonnet 5 is the only exception: MOpenHands has a higher inference cost (\$0.6222 vs. \$0.5634/instance) but also a higher success rate (57.94\% vs. 31.18\%).
These results show that method selection affects both reproduction test generation performance and inference cost.

\section{Related Work}
\label{sec:related-work}
% In this section, we review related work on code-related benchmarks for LLMs and reproduction test generation benchmarks.

\textbf{Code-Related Benchmarks for LLMs.}
A wide range of benchmarks has been developed to evaluate the capabilities and limitations of LLMs on code-related tasks.
Early benchmarks primarily focused on program-level tasks in a single programming language~\citep{allamanis2013msr,raychev2016sigplan,iyer2018emnlp,chen2021arxiv,austin2021arxiv,wang2023emnlp}.
As LLMs have advanced, code-related benchmarks have increasingly moved toward evaluation settings that better reflect real-world software engineering, particularly along two dimensions: \emph{multilingual evaluation} and \emph{repository-level evaluation}.
% First, several studies have extended existing code-generation benchmarks beyond a single programming language.
For multilingual evaluation, Multilingual-HumanEval~\citep{athiwaratkun2023iclr} and HumanEval-X~\citep{zheng2023kdd} extend HumanEval~\citep{chen2021arxiv} to multiple programming languages, while MBXP~\citep{athiwaratkun2023iclr} provides a multilingual extension of MBPP~\citep{austin2021arxiv}.
% These benchmarks enable us to examine whether model capabilities observed in one programming language generalize across different language ecosystems.
% In parallel, benchmarks have shifted from program-level tasks toward repository-level software engineering tasks, such as repository-level code completion~\citep{zhang2023emnlp,liu2024iclr,ding2023nips}, bug fixing~\citep{mundler2024nips,ouyang2024issta,saavedra2024icse}, and unit test generation~\citep{quang2026arxiv}.
% These benchmarks require models to reason about broader code contexts, dependencies, and interactions across multiple files, thereby providing a more realistic evaluation of their ability to handle software engineering tasks.
To evaluate LLMs in more complex scenarios, benchmarks have expanded from program-level tasks to repository-level tasks that require reasoning about dependencies and interactions across multiple files, including code completion~\citep{zhang2023emnlp,liu2024iclr,ding2023nips}, bug fixing~\citep{mundler2024nips,ouyang2024issta,saavedra2024icse}, and unit test generation~\citep{quang2026arxiv}.

In particular, \swe~\citep{jimenez2024iclr} has become a prominent benchmark for repository-level issue resolution, evaluating LLMs on real-world GitHub issues by requiring them to modify a repository to resolve a given issue.
It has driven substantial progress in evaluating and improving LLM-based issue resolution~\citep{deng2025arxiv,huang2026arxiv}.
Building on this setting, \multiswe~\citep{zan2026nips} extends repository-level issue resolution to eight programming languages, enabling systematic evaluation across diverse software ecosystems.
% Together, these efforts reflect a broader shift toward evaluating LLMs on realistic repository-level tasks across multiple programming languages.

% Building on \swe~\citep{jimenez2024iclr}, multilingual benchmarks have further expanded repository-level issue resolution across multiple programming languages.
% \multiswe~\citep{zan2026nips} covers seven programming languages, including Java, TypeScript, JavaScript, Go, Rust, C, and C++, enabling systematic evaluation of LLMs across diverse software ecosystems.
% Together, these efforts reflect a broader shift toward evaluating LLMs across diverse programming languages and repository-level software engineering tasks.

\textbf{Benchmarks for Reproduction Test Generation.}
\swt~\citep{mundler2024nips} and TDD-Bench-Verified~\citep{ahmed2024arxiv} evaluate LLMs on reproduction test generation in Python.
Both are derived from the \swe dataset~\citep{jimenez2024iclr} and assess generated tests by checking whether they fail on the original code and pass after the issue has been resolved.
% Although they share the same task and evaluation principle, they differ in dataset construction and evaluation details.
% TDD-Bench-Verified focuses on carefully filtered instances with validated contributing tests, whereas \swt provides broader settings and variations, including \swelite and \sweverified subsets.
Beyond Python, TDD-Bench-Java~\citep{ahmed2026arxiv} extends repository-level reproduction test generation to Java with 250 instances from popular open-source repositories, demonstrating the need to adapt existing approaches to the Java ecosystem.
% This study further shows that reproduction test generation can involve language-specific challenges, such as differences in compilation, test organization, and repository structure, requiring adaptations to approaches originally developed for Python.
% These findings highlight the importance of evaluating reproduction test generation across programming languages rather than assuming that performance and design choices observed in one language generalize to others.
% This work demonstrates the importance of evaluating reproduction test generation beyond Python, since programming languages differ in their type systems, build processes, testing frameworks, and repository structures.
Although these benchmarks broaden language coverage, their independently constructed monolingual settings do not provide a unified view of how the same agents perform across programming ecosystems.
Our dataset complements them by applying a common task definition, validation criterion, and evaluation protocol to tasks from eight programming languages, enabling consistent evaluation across language-specific subsets.
% This setting enables consistent evaluation of the same agents across language-specific subsets and reveals performance variations that may be overlooked by single-language evaluations.

\section{Conclusion and Future Work}
\label{sec:conclusion}

In this study, we introduce \multiswt, a multilingual benchmark for reproduction test generation consisting of 1,963 instances across eight programming languages.
Using \multiswt, we evaluate state-of-the-art LLMs with four representative reproduction test generation methods.
Based on our empirical results and failure analysis, we distill lessons for evaluating and developing automated approaches and discuss promising opportunities for future research.

\begin{itemize}
    \item \textbf{Multilingual evaluation}: Python is the only language whose success rate consistently exceeds the aggregate success rate across all evaluated configurations (Section~\ref{sec:performance-languages}), indicating that performance measured only on Python may not transfer to ecosystems with different build systems and testing practices.
    Evaluations should therefore cover diverse programming ecosystems to assess the generalizability.    
    \item \textbf{Ecosystem-aware methods}: Our failure analysis shows that valid reproduction tests depend on ecosystem-specific conventions (Section~\ref{sec:language-specific-challenges}). Agents should inspect how tests are registered and aggregated and run repository-level checks after adding a test.
    \item \textbf{Behavioral validation}: Although MOpenHands, Codex, and Claude Code generate failing tests for over 80\% of instances, many failures are not resolved by the golden patch (Section~\ref{sec:performance-methods-llms}). Our failure analysis highlights two recurring challenges: (1) the required setup may be implicit, and (2) iterative revisions can remove the target behavior or make a previously valid test non-executable (Section~\ref{sec:cross-language-challenges}). Future methods should therefore identify the expected behavioral difference between buggy and fixed versions, infer the necessary setup from the issue and repository context, and revalidate it after each revision.
\end{itemize}

\subsection*{AI use statement}
In this work, we used generative AI tools for the following tasks with required disclosure: design or provide feedback on research methodology or experiments; implement methods; assist with translation; polish writing; identify relevant literature; and clean and reformat dataset.
We have not used generative AI tools for the following tasks with required disclosure: generate synthetic data sets; propose or refine hypotheses; interpret results; and support qualitative and thematic data analysis.
The remaining tasks with required disclosure are not applicable to this work: help develop theoretical models or conceptual frameworks; formulate mathematical claims; provide critical ingredients for proving mathematical claims; and assist in the writing of proofs.
We have reviewed all AI-assisted work: references identified with AI assistance were checked against their original sources, and all numerical results and claims were verified against the original experimental results.
We take responsibility for the final content of this work, including text, claims, and artifacts produced with the aid of generative AI.
% (This section is \textbf{required} and does not count toward the page limit.)

% In this work, we used generative AI tools for [tasks with required disclosure].
% We have not used generative AI tools for [other tasks with required disclosure],
% and [the rest of the required disclosure tasks] are not applicable to this work.
% Additionally, we used generative AI tools for [tasks with recommended
% disclosure]. We have reviewed all AI-assisted work. [Elaborate. For example, ``we
% checked LLM-generated research ideas for potential plagiarism through a manual
% literature survey'', ``LLM-generated code was verified and tested for correctness
% by 2 authors'', etc.]. We take responsibility for the final content of this work,
% including text, claims or artifacts produced with the aid of generative AI.

% See the ICLR 2027 AI Policy for Authors for more details. This statement should
% not be more than 1 page.

\subsection*{Reproducibility statement}
To support reproducibility, we provide a replication package for this study.
It contains the benchmark data, evaluation scripts, experimental configurations, prompts, and instructions for reproducing the main results.
Section~\ref{sec:multiswt} describes the construction and validation of \multiswt.
Section~\ref{sec:experimental-setup} describes the experimental setup and evaluation metrics.
The prompts are also provided in Appendix~\ref{app:prompts}.
The replication package is available at: \url{https://doi.org/10.5281/zenodo.22961970}.

% 本研究の再現性を確保するため，再現用パッケージを提供する．このパッケージには，評価基盤のデータ，評価用スクリプト，実験設定，指示文，主要な結果を再現するための手順が含まれる．第3章では，本評価基盤の構築と検証方法を説明している．第4章では，実験設定と評価指標を説明している．指示文は付録Aにも掲載している．再現用パッケージは次の場所で公開している：[匿名URL]．

% (This section is \textbf{recommended} and does not count toward the page limit.)

% It is important that the work published in ICLR is reproducible. Authors are
% strongly encouraged to include a paragraph-long Reproducibility Statement at the
% end of the main text (before references) to discuss the efforts that have been
% made to ensure reproducibility. This paragraph should not itself describe
% details needed for reproducing the results, but rather reference the parts of
% the main paper, appendix, and supplemental materials that will help with
% reproducibility. For example, for novel models or algorithms, a link to an
% anonymous downloadable source code can be submitted as supplementary materials;
% for theoretical results, clear explanations of any assumptions and a complete
% proof of the claims can be included in the appendix; for any datasets used in
% the experiments, a complete description of the data processing steps can be
% provided in the supplementary materials. Each of the above are examples of
% things that can be referenced in the reproducibility statement.

% \input{section/ethics}  % Uncomment if ethics concerns apply

\bibliography{references}
\bibliographystyle{template/iclr2027_bibliography}

\appendix
\raggedbottom
\section{Prompts} \label{app:prompts}

The prompts used for MSWE-agent (Figure~\ref{fig:mswe-agent-prompt}) and MOpenHands (Figure~\ref{fig:mopenhands-prompt}), as well as the prompt used for Codex and Claude Code (Figure~\ref{fig:codex-claude-prompt}), are shown below.
Key changes for the reproduction test generation task are highlighted in \textbf{boldface}.

\begin{figure}[p]
\centering
\begin{lstlisting}[
breaklines=true,
language=,
frame=single,
escapeinside={(*@}{@*)}
]
We have received following issue within our repository. Here's the issue text:
ISSUE:
(*@\textit{user issue comes here}@*)

INSTRUCTIONS:
Now, you're going to create unit tests that cover the issue. In other words, you should write unit tests that fail in the current state of the repository but will pass when the issue has been resolved. Essentially, you'll want to write a unit test that reproduces the described issue.
Your terminal session has started and you're in the repository's root directory. You can use any bash commands or the special interface to help you. Edit all the files you need to and run any checks or tests that you want.
Remember, YOU CAN ONLY ENTER ONE COMMAND AT A TIME. You should always wait for feedback after every command.
When you're satisfied with all of the changes you've made, you can submit your changes to the code base by simply running the submit command.
Note however that you cannot use any interactive session commands (e.g. an interactive interpreter, vim) in this environment, but you can write scripts and run them. E.g. you can write a (*@\textit{target language}@*) script and then run it with `(*@\textit{language-specific script command}@*)`.

NOTE ABOUT THE EDIT COMMAND: Indentation really matters! When editing a file, make sure to insert appropriate indentation before each line!

IMPORTANT TIPS:
1. Always start by trying to replicate the bug that the issues discusses.
   If the issue includes code for reproducing the bug, we recommend that you re-implement that in your environment, and run it to make sure you can reproduce the bug.
   Then start trying to fix it.
   When you think you've fixed the bug, re-run the bug reproduction script to make sure that the bug has indeed been fixed.

   If the bug reproduction script does not print anything when it successfully runs, we recommend adding a statement that prints "Script completed successfully, no errors." in (*@\textit{target language}@*) at the end of the file,
   so that you can be sure that the script indeed ran fine all the way through.

2. If you run a command and it doesn't work, try running a different command. A command that did not work once will not work the second time unless you modify it!

3. If you open a file and need to get to an area around a specific line that is not in the first 100 lines, say line 583, don't just use the scroll_down command multiple times. Instead, use the goto 583 command. It's much quicker.

4. If the bug reproduction script requires inputting/reading a specific file, such as buggy-input.png, and you'd like to understand how to input that file, conduct a search in the existing repo code, to see whether someone else has already done that. Do this by running the command: find_file "buggy-input.png" If that doesn't work, use the linux 'find' command.

5. Always make sure to look at the currently open file and the current working directory (which appears right after the currently open file). The currently open file might be in a different directory than the working directory! Note that some commands, such as 'create', open files, so they might change the current open file.

6. When editing files, it is easy to accidentally specify a wrong line number or to write code with incorrect indentation. Always check the code after you issue an edit to make sure that it reflects what you wanted to accomplish. If it didn't, issue another command to fix it.


\end{lstlisting}
\caption{The Prompt for MSWE-agent on \multiswt}
\label{fig:mswe-agent-prompt}
\end{figure}

\begin{figure}[p]
    \centering
    \begin{lstlisting}[
        breaklines=true,
        language=,
        frame=single,
        escapeinside={(*@}{@*)}
    ]
<uploaded_files>
(*@\textit{workspace path comes here}@*)
</uploaded_files>

I've uploaded a (*@\textit{target language}@*) code repository in the directory (*@\textit{workspace directory}@*). Consider the following issue description:

<issue_description>
(*@\textit{user issue comes here}@*)
</issue_description>

Can you help me implement the necessary changes to the repository to (*@\textbf{test whether the issue in <issue\_description> was resolved}@*)?

I will take care of all changes to any non-test files. This means you DON'T have to modify the actual implementation logic. (*@\textbf{ONLY update test logic and tests. Do not fix the issue itself.}@*)

The development (*@\textit{target language}@*) environment is already set up for you, so you don't need to install other packages.

Your task is to make the minimal changes to test files in the /workspace directory to (*@\textbf{reproduce the issue in the <issue\_description>. In other words, you should write unit tests that fail in the current state of the repository but will pass when the issue has been resolved.}@*)

Follow these steps to reproduce the issue:
1. Explore the repository to understand its structure and test framework.
2. Create a script, class, or executable to reproduce the error and run it using the appropriate command for the target programming language to confirm the error.
3. Integrate the reproduction into the repository's existing test framework by editing only test files.
4. Before finishing, check the diff against the base commit (*@\textit{base commit comes here}@*). Remove standalone reproduction scripts, build outputs, dependency changes, and other temporary artifacts that are not part of the intended test patch.

Your thinking should be thorough and so it's fine if it's very long.
    \end{lstlisting}
    \caption{The Prompt for MOpenHands on \multiswt}
    \label{fig:mopenhands-prompt}
\end{figure}
\begin{figure}[p]
    \centering
    \begin{lstlisting}[
        breaklines=true,
        language=,
        frame=single,
        escapeinside={(*@}{@*)}
    ]
We have received following issue within our repository. Here's the issue text:

ISSUE:
(*@\textit{user issue comes here}@*)

INSTRUCTIONS:
Now, you're going to create unit tests that cover the issue. In other words, you should write unit tests that fail in the current state of the repository but will pass when the issue has been resolved. Essentially, you'll want to write a unit test that reproduces the described issue.

Your terminal session has started and you're in the repository's root directory. You can use any bash commands to help you. Edit all the files you need to and run any checks or tests that you want.
    \end{lstlisting}
    \caption{The Prompt for Codex and Claude Code on \multiswt}
    \label{fig:codex-claude-prompt}
\end{figure}

\section{Additional Analyses of Factors Influencing Performance}
\label{app:influencing-factors}
To identify factors associated with reproduction test generation performance, we focus on four key factors: (1) \textit{issue type}, (2) \textit{issue description characteristics}, (3) \textit{fix patch characteristics}, and (4) \textit{reference test characteristics}.

\subsection{Issue Type}
\label{app:issue-type}
We analyze whether reproduction test generation performance differs across three issue types: bug fixes (Bug Fix), new features (New Feat.), and feature optimizations (Feat. Opt.).
% This analysis examines whether the type of behavioral change described in an issue affects the difficulty of constructing a test that distinguishes the original and issue-resolving versions.
We use the manually assigned issue-type labels provided by \multiswe~\citep{zan2026nips}.

Table~\ref{tab:joint-issue-type-1} reports reproduction test generation performance across issue types and programming languages.
We do not observe a consistent performance hierarchy across issue types.
Although bug fix issues achieve the highest success rates in several settings, this trend does not hold across all languages and methods.
In Go, MSWE-agent achieves a lower success rate for bug fixes (21.05\%) than for new features (40.00\%) and feature optimizations (50.00\%), whereas MOpenHands and Codex show the opposite trend, achieving their highest success rates for bug fixes.
% % Similarly, for JavaScript feature optimization issues, MSWE-agent and MOpenHands both achieve 100.00\%, while Codex achieves 0.00\%.
% This result indicates that the difficulty of generating reproduction tests for different issue types depends not only on the type of behavioral change, but also on the programming language and the agent used.

\begin{table}[H]
\centering
\renewcommand{\arraystretch}{1.1}
\caption{Success rate (\%) across issue types and languages (GPT-5.6 Luna).}
\label{tab:joint-issue-type-1}
\small
\setlength{\tabcolsep}{2.5pt}
\begin{tabular}{lccccccccc}
\toprule
Language & \multicolumn{3}{c}{\textbf{MSWE-agent}} & \multicolumn{3}{c}{\textbf{MOpenHands}} & \multicolumn{3}{c}{\textbf{Codex}} \\
\cmidrule(lr){2-4} \cmidrule(lr){5-7} \cmidrule(lr){8-10}
 & Bug Fix & New Feat. & Feat. Opt. & Bug Fix & New Feat. & Feat. Opt. & Bug Fix & New Feat. & Feat. Opt. \\
\midrule
% Python & -- & -- & -- & -- & -- & -- & -- & -- & -- \\
Java & 61.11 & 0.00 & 0.00 & 50.00 & 33.33 & 100.00 & 41.67 & 33.33 & 0.00 \\
TS & 35.56 & -- & -- & 66.67 & -- & -- & 55.56 & -- & -- \\
JS & 48.00 & 26.32 & 100.00 & 48.00 & 36.84 & 100.00 & 52.00 & 31.58 & 0.00 \\
Go & 21.05 & 40.00 & 50.00 & 42.11 & 25.00 & 33.33 & 47.37 & 15.00 & 33.33 \\
Rust & 60.87 & 45.00 & 50.00 & 52.17 & 35.00 & 50.00 & 47.83 & 35.00 & 50.00 \\
C & 32.35 & 25.00 & 0.00 & 47.06 & 50.00 & 50.00 & 44.12 & 25.00 & 50.00 \\
C++ & 25.00 & 28.57 & 0.00 & 37.50 & 35.71 & 0.00 & 33.33 & 35.71 & 0.00 \\
\bottomrule
\end{tabular}
\end{table}

\subsection{Characteristics of Issue Description}
\label{app:issue-description-characteristics}
\begin{figure}[H]
\centering
\includegraphics[width=0.6\linewidth]{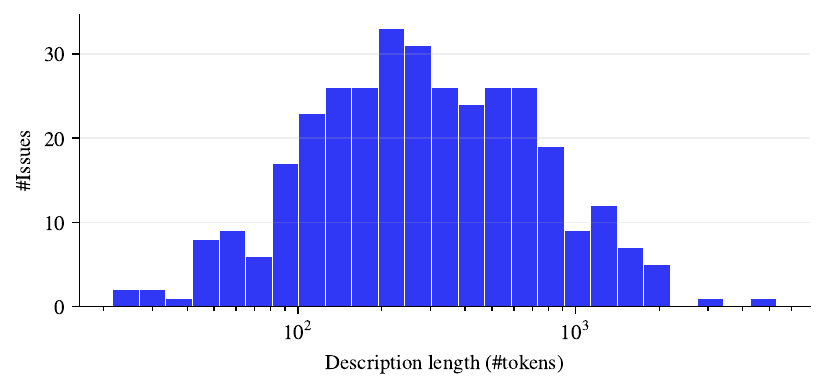}
\caption{Distribution of issue description length.}
\label{fig:description-figure9}
\end{figure}

We investigate how issue description length affects reproduction test generation performance.
Figure~\ref{fig:description-figure9} shows the distribution of issue description lengths in \multiswt, with the majority of issues containing fewer than 1,000 tokens.
% Most issues have relatively short descriptions, while only a small number have very long descriptions.
To examine the effect of issue description length, we divide the issues into five intervals: $<$100, 100–400, 400–700, 700–1,000, and $>$1,000 tokens.
% Figure~\ref{fig:issue-description-length} reports the mean success rate of each method for each interval across programming languages.

\begin{figure}[H]
\centering
\includegraphics[width=0.95\linewidth]{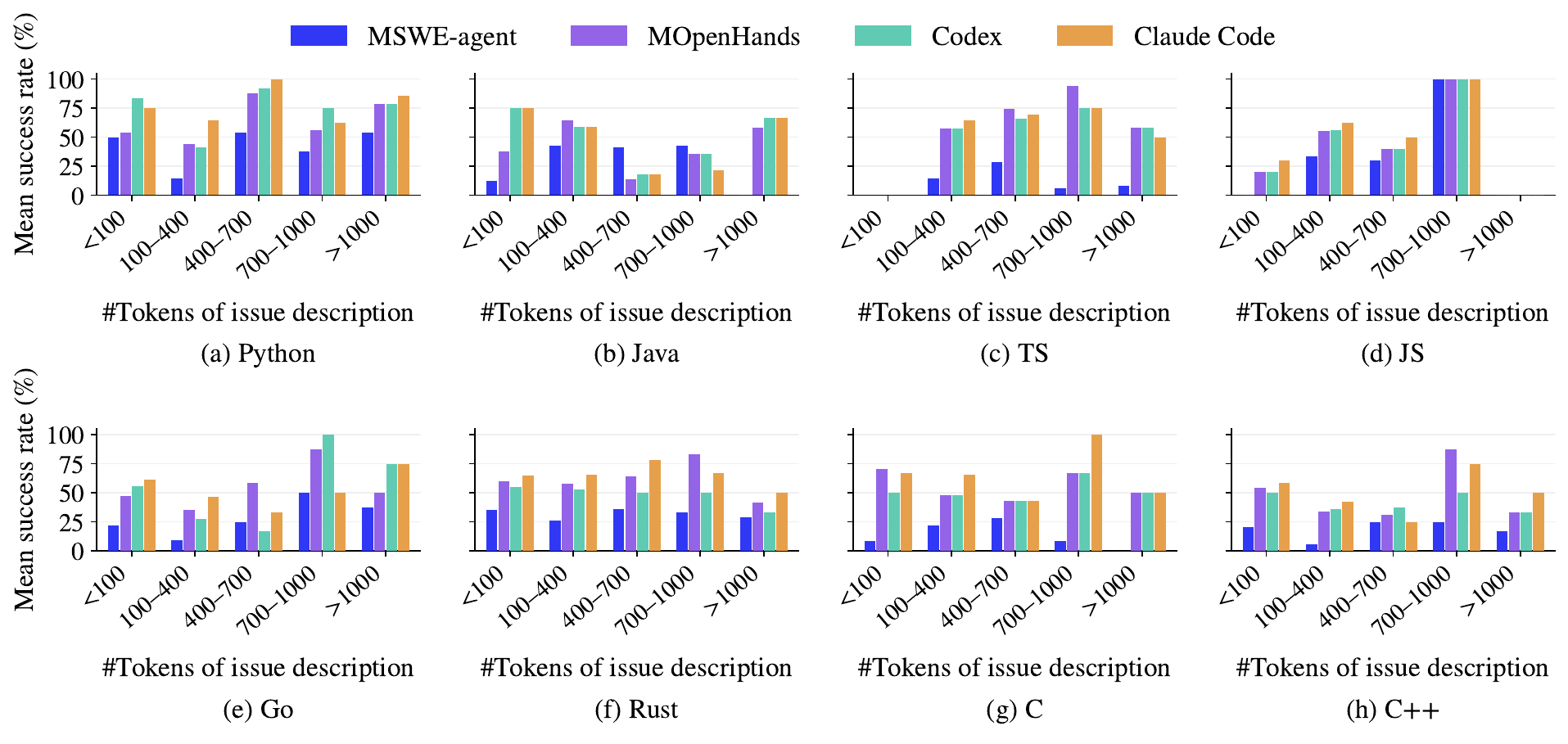}
\caption{Mean success rate by issue description length.}
\label{fig:issue-description-length}
\end{figure}

As shown in Figure~\ref{fig:issue-description-length}, there is no consistent relationship between issue description length and reproduction test generation performance.
Longer descriptions may correspond to two qualitatively different types of issues: (1) detailed issue reports that provide precise indications of relevant code locations and concrete steps for resolving the issue, which facilitate reproduction test generation, and (2) intrinsically complex issues that require lengthy descriptions to explain their behavior, which make reproduction test generation more difficult.
% The trend varies across programming languages.
% In Python, all four agents achieve higher success rates for issues with 400--700 tokens than for those with 100--400 tokens, whereas in Java, all four agents show the opposite trend.
% The relationship also varies across agents within the same language.
% For example, in TypeScript, MOpenHands achieves a high success rate for issues with 700--1,000 tokens, while MSWE-agent shows a substantially lower success rate for the same interval.
% These results indicate that the relationship between issue description length and reproduction test generation performance varies depending on the programming language and agent.

% \newpage

\subsection{Characteristics of Golden Fix Patches}
\label{app:golden-fix-patch-characteristics}

In this subsection, we investigate the relationship between golden fix patch characteristics and reproduction test generation performance, focusing on two factors:
(1) \textit{fix patch length}: We examine how the size of the issue-resolving changes is associated with reproduction test generation performance.
% Figure~\ref{fig:golden-fix-s523_length_dist} shows the distribution of fix patch lengths, measured by tokenizing the complete fix diff with \texttt{o200k\_base}.
Figure~\ref{fig:golden-fix-s523_length_dist} shows the distribution of fix patch lengths, measured as the number of tokens in the complete fix diff.
We divide the patches into five intervals: $<$200, 200--600, 600--1,000, 1,000--1,400, and $>$1,400 tokens.
(2) \textit{number of modified files}: We examine whether reproduction test generation performance differs between instances whose golden fix patch modifies a single file and those whose golden fix patch modifies multiple files.
Figure~\ref{fig:golden-fix-s523_files_dist} shows the distribution of the number of files modified by each golden fix patch.
We divide the instances into two groups: single-file fixes, whose golden fix patches modify exactly one file, and multi-file fixes, whose golden fix patches modify two or more files.
% Figure~\ref{fig:golden-fix-s523_files_dist} shows the distribution of the number of modified files, which we divide into four groups: 1, 2--5, 6--10, and $>$10 files.
% The figures report the mean success rate across the evaluated LLMs within each agent.

% \paragraph{Performance generally decreases as fix patch length increases.}
% Figure~\ref{fig:golden-fix-s523_length} shows the relationship between golden fix patch length and reproduction test generation success rate.
% In Python, TypeScript, JavaScript, Rust, and C++, all four agents achieve higher success rates for patches shorter than 200 tokens than for patches longer than 1,400 tokens.
% The difference is particularly pronounced in JavaScript, where the success rate decreases from 75.00\% to 9.62\% for MSWE-agent, from 83.33\% to 32.69\% for MOpenHands, from 91.67\% to 30.77\% for Codex, and from 91.67\% to 42.31\% for Claude Code.
% These results suggest that issues involving larger fixes tend to be more difficult to reproduce, even though reproduction test generation does not require the agent to generate the fix itself. One possible explanation is that larger fixes reflect behavioral changes involving a broader portion of the codebase, making it more difficult to identify the behavior that the generated test should capture.
% However, this trend is not uniform across all languages, as C shows higher success rates for the longest-patch group than for the shortest-patch group.

\begin{figure}[H]
    \centering
    \input{figure/results/s52/s523_length_dist.tex}%
    \hfill
    \input{figure/results/s52/s523_files_dist.tex}
\end{figure}

\begin{figure}[H]
\centering
\includegraphics[width=0.95\linewidth]{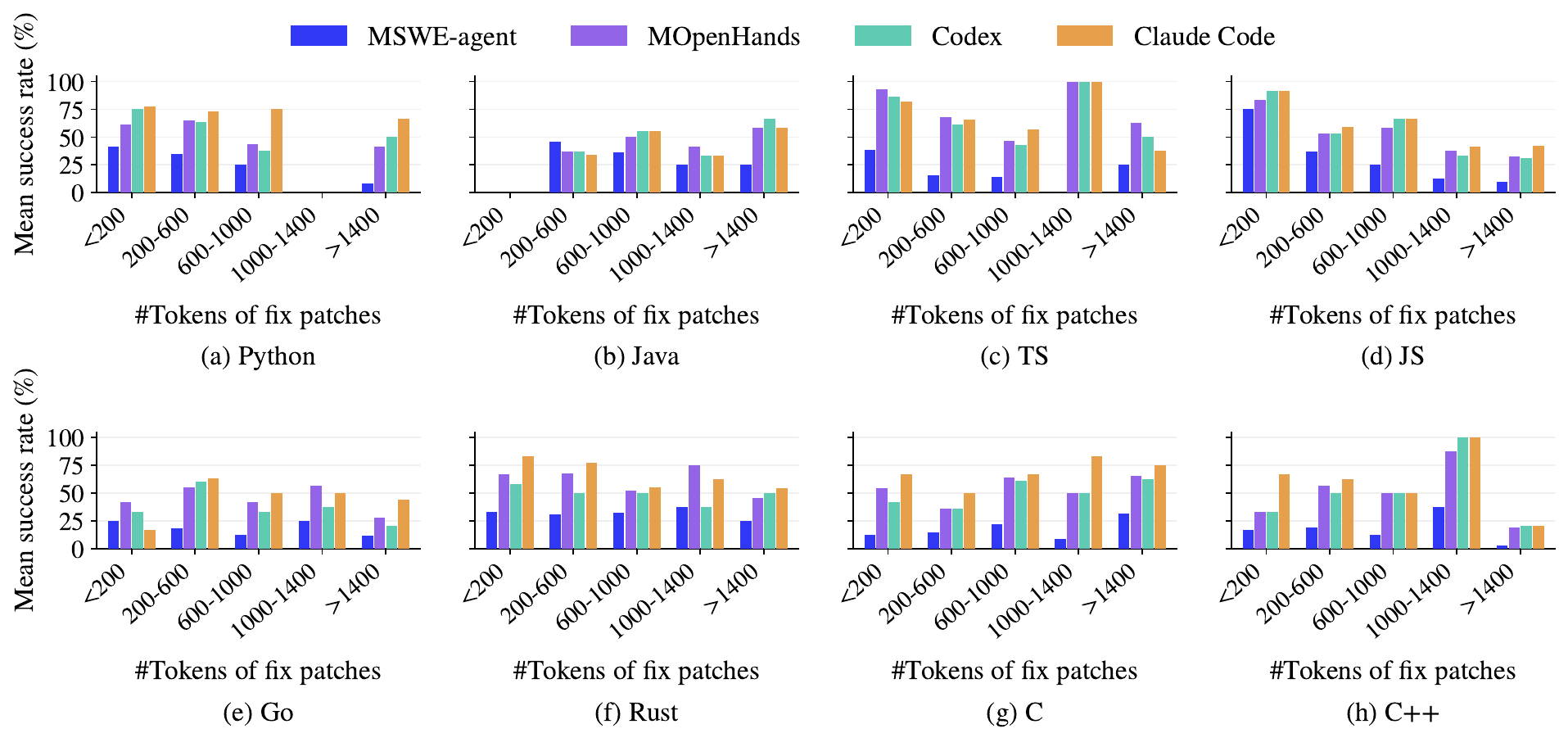}
\caption{Mean success rate by golden fix patch length.}
\label{fig:golden-fix-s523_length}
\end{figure}

As shown in Figure~\ref{fig:golden-fix-s523_length}, there is no consistent relationship between fix patch length and reproduction test generation performance.
This may be because fix patch length does not necessarily reflect the complexity of the behavior that a reproduction test must capture.
Longer patches can reflect broad or complex behavioral changes, but they can also consist of repetitive or mechanical changes, while short patches can address subtle behavior that is difficult to trigger.

% \paragraph{Cross-file fix patches are associated with reduced effectiveness.}
% Figure~\ref{fig:golden-fix-s523_files} illustrates the relationship between the number of files modified by the golden fix and reproduction test generation success rate.
% In Java, JavaScript, Go, and C++, all four agents achieve higher success rates for single-file fixes than for fixes spanning 2--5 files.
% For example, in Java, the success rate decreases from 56.82\% to 31.25\% for MSWE-agent and from 56.82\% to 40.18\% for MOpenHands.
% These results suggest that reproduction test generation becomes more challenging when the issue-resolving changes span multiple files.
% Such issues may involve interactions across different code regions, requiring the agent to understand a broader portion of the repository to identify the behavior that should be reproduced.
% Although exceptions exist, such as Rust for MSWE-agent and Codex, the overall results highlight cross-file understanding as an important challenge for reproduction test generation.

\begin{figure}[H]
\centering
\includegraphics[width=0.8\linewidth]{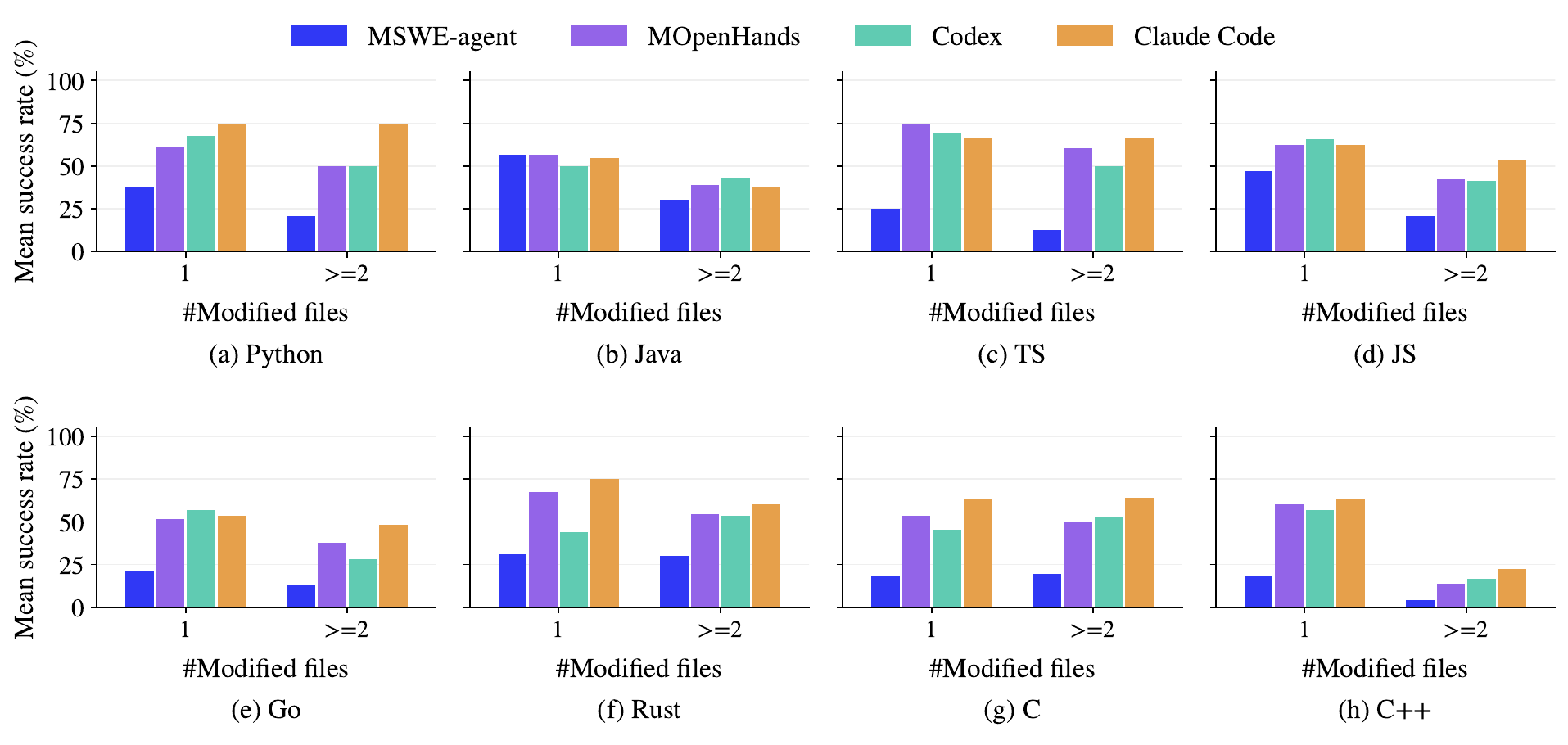}
\caption{Mean success rate by the number of files modified by golden fixes.}
\label{fig:golden-fix-s523_files_binary}
\end{figure}

As shown in Figure~\ref{fig:golden-fix-s523_files_binary}, reproduction test generation performance is lower for multi-file fixes than for single-file fixes in most programming languages.
This trend is observed for most agents, with only a limited number of cases in which multi-file fixes achieve higher success rates.
These results suggest that issues whose golden fixes span multiple files tend to be more difficult to reproduce.
% ファイル数を1, 2--5, 6--10, and $>$10で分けてた時の文章
% Figure~\ref{fig:golden-fix-s523_files} also shows no consistent relationship between the number of files modified by the golden fix and reproduction test generation performance.
% Cross-file fixes can involve complex interactions across code regions, but they can also consist of simple, repetitive changes across files.
% Thus, modifying multiple files does not necessarily make the issue more difficult to reproduce.

% \input{figure/results/s52/s523_files.tex}

\subsection{Characteristics of Golden Reference Tests}

\label{app:gold-test-characteristics}

In this subsection, we investigate the relationship between the characteristics of golden reference tests and reproduction test generation performance, focusing on two factors: (1) \textit{golden reference test length} and (2) \textit{number of executed files}.
% For golden reference test length, we examine whether the size of the tests required to reproduce an issue is associated with reproduction test generation performance.
Figure~\ref{fig:gt-length-dist} shows the distribution of golden reference test lengths, measured as the total number of tokens in the developer-written reference tests for each instance.
Based on this distribution, we divide the tests into four intervals: $\leq 300$, 301--600, 601--900, and $\geq 901$ tokens.
% For execution scope, we examine how broadly the golden reference tests exercise the repository during execution.
% We measure execution scope as the number of distinct files executed when running the golden reference tests.
% This metric differs from the number of files modified by the golden fix.
% The number of modified files represents the scope of the issue-resolving changes, whereas the number of executed files represents the runtime scope required to exercise the behavior targeted by the reproduction test.
Figure~\ref{fig:gt-files-dist} shows the distribution of the number of files executed when running the golden reference tests.
We divide the instances into two groups: single-file executions, in which the golden reference tests execute exactly one file, and multi-file executions, in which they execute two or more files.
% The number of executed files spans a wide range, including cases in which hundreds of files are executed.

\begin{figure}[H]
    \centering
    \input{figure/results/s52/sub/gt_length_dist.tex}
    \hfill
    \input{figure/results/s52/sub/gt_files_dist.tex}
\end{figure}

\begin{figure}[H]
\centering
\includegraphics[width=0.95\linewidth]{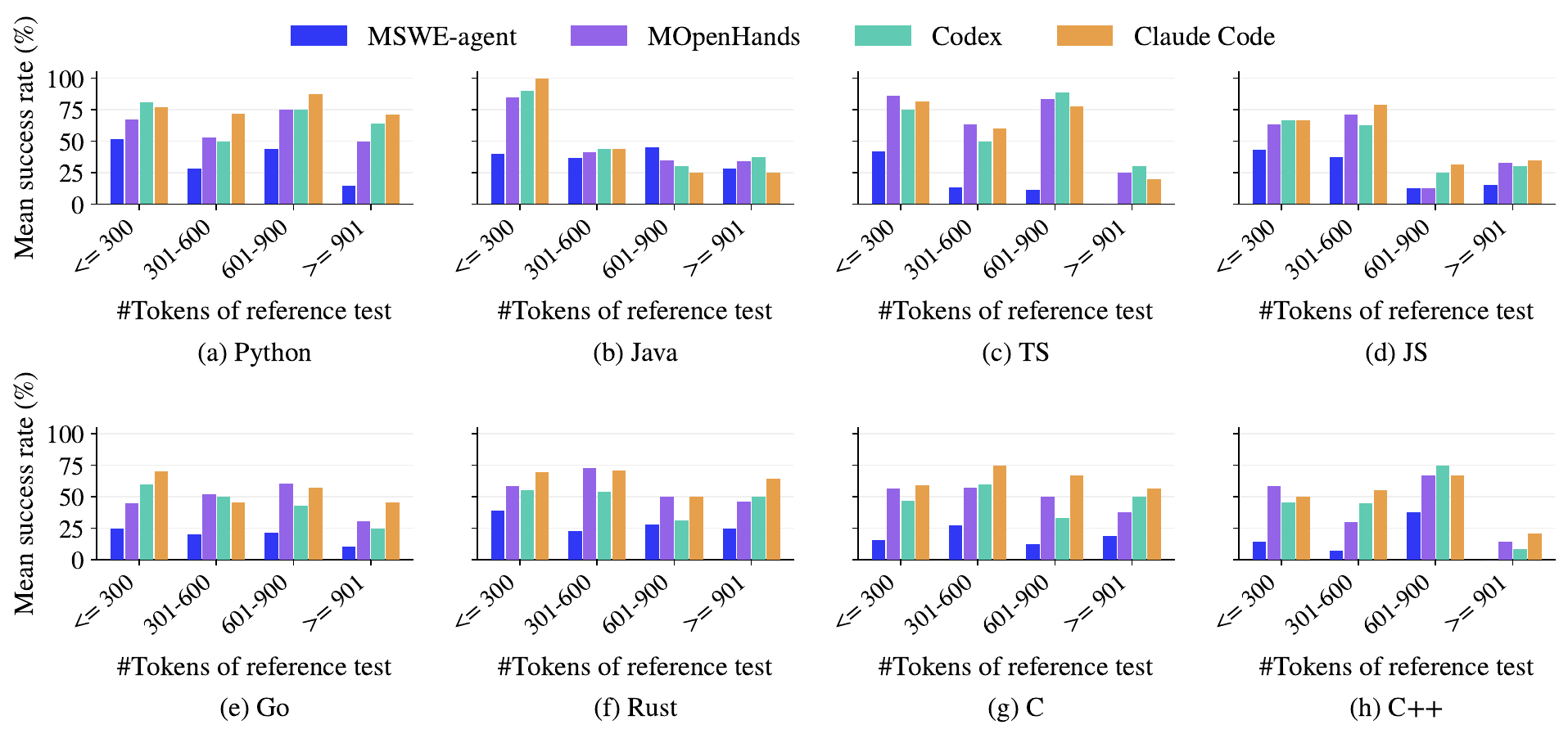}
\caption{Mean success rate by golden reference test length.}
\label{fig:gt-length}
\end{figure}

As shown in Figure~\ref{fig:gt-length}, there is no monotonic relationship between golden reference test length and reproduction test generation performance.
The success rate does not consistently decrease as the golden reference tests become longer, and intermediate-length groups sometimes achieve higher success rates than shorter groups.
However, the $\geq 901$-token group generally shows lower success rates than the $\leq 300$-token group across most programming languages and agents.
These results suggest that issues with relatively long golden reference tests tend to be more difficult to reproduce, although golden reference test length alone does not consistently explain reproduction test generation performance.

\begin{figure}[H]
\centering
\includegraphics[width=0.8\linewidth]{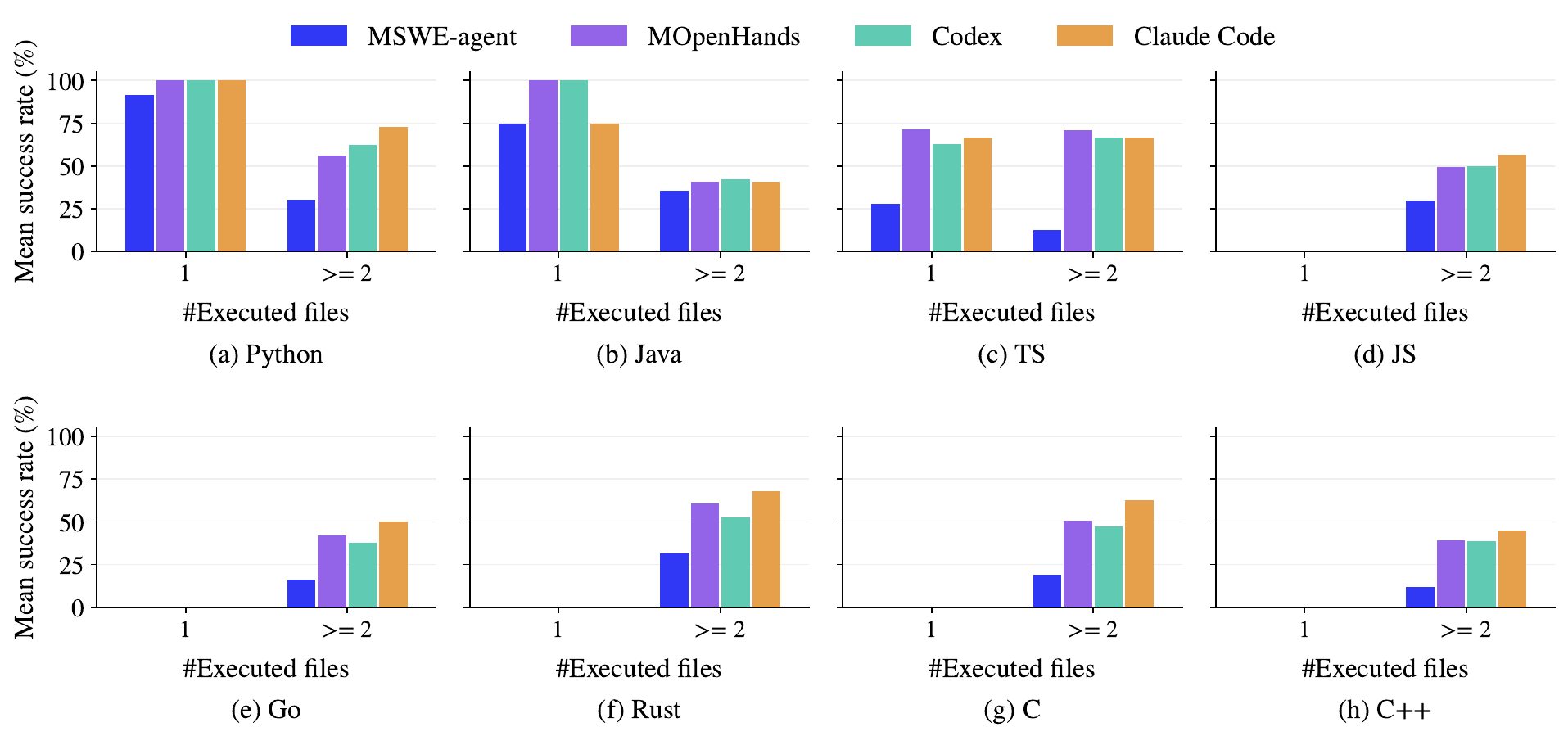}
\caption{Mean success rate by the number of files executed by golden reference tests.}
\label{fig:gt-files}
\end{figure}

As shown in Figure~\ref{fig:gt-files}, the relationship between execution scope and reproduction test generation performance varies across programming languages.
For Python and Java, tests that execute multiple files show lower success rates than those that execute a single file across all evaluated agents.
In contrast, this pattern is not consistently observed for TypeScript.
For the other programming languages, a meaningful comparison between single-file and multi-file execution is difficult because the single-file group is not sufficiently represented.
These results indicate that the relationship between execution scope and reproduction test generation performance is strongly dependent on the programming language, and no consistent language-general trend is observed.

% \newpage

\section{Detailed Examples from the Failure Analysis}
\label{app:failure-examples}

\subsection{Language-Specific Challenges}
\subsubsection{Testing Conventions}
\label{app:failure-testing-conventions}

\paragraph{ripgrep \#1642.}
\label{app:failure-ripgrep-1642}

\href{https://github.com/BurntSushi/ripgrep/issues/1380}{\texttt{ripgrep \#1380}}
reports that combining a match limit with trailing context can produce more output than requested.
The corresponding benchmark instance is derived from
\href{https://github.com/BurntSushi/ripgrep/pull/1642}{\texttt{PR \#1642}}.
The agent generated a test using \texttt{-m 1 -A 2} and expected the first match followed by two context lines.

Listing~\ref{lst:ripgrep-1642-test} shows the generated test.
Although the expected output contains line-number prefixes, the generated command does not enable \texttt{-n} or \texttt{--line-number}.
As shown in Listing~\ref{lst:ripgrep-1642-output}, before the golden patch, the command produces too many lines, reproducing the original issue.
After the golden patch, the output is correctly reduced to three lines, but the lines still do not contain the prefixes expected by the generated test.
Consequently, the test fails both before and after the golden patch.

\begin{minipage}{\linewidth}
\begin{lstlisting}[
    caption={Generated reproduction test for ripgrep PR \#1642.},
    label={lst:ripgrep-1642-test}
]
rgtest!(max_count_after_context_match, |dir: Dir, mut cmd: TestCommand| {
    dir.create("rgtest.txt", "a\nb\nc\nd\ne\nd\ne\nd\ne\nd\ne\n");
    cmd.args(&["-m", "1", "-A", "2", "d", "rgtest.txt"]);

    let expected = "4:d\n5-e\n6:d\n";
    eqnice!(expected, cmd.stdout());
});
\end{lstlisting}
\end{minipage}

\begin{minipage}{\linewidth}
\begin{lstlisting}[
    language=diff,
    caption={Observed output before and after applying the golden patch.},
    label={lst:ripgrep-1642-output}
]
Expected:
  4:d
  5-e
  6:d

Before golden patch:
- d
- e
- d
- e
- d
- e
- d
- e

After golden patch:
+ d
+ e
+ d
\end{lstlisting}
\end{minipage}

\paragraph{Jackson Databind \#1923.}
\label{app:failure-jackson-1923}

\href{https://github.com/FasterXML/jackson-databind/issues/1872}{\texttt{Jackson Databind \#1872}}
reports a null-pointer exception when deserializing a collection whose element type is a Spring-related interface.
The corresponding benchmark instance is derived from
\href{https://github.com/FasterXML/jackson-databind/pull/1923}{\texttt{PR \#1923}}.
The golden patch prevents subtype validation from traversing the superclass of an interface, thereby eliminating the null-pointer failure.

Listing~\ref{lst:jackson-1923-test} shows the generated test.
Rather than directly checking whether the interface can be handled without the reported null-pointer exception, the test reuses an existing helper for prohibited types.
As shown in Listing~\ref{lst:jackson-1923-helper}, this helper requires the resulting exception message to contain security-specific phrases such as \texttt{Illegal type} and \texttt{prevented for security reasons}.
After applying the golden patch, the null-pointer exception is eliminated, but deserialization instead produces an ordinary error because the interface cannot be instantiated.
Since this error does not contain the security-specific message required by the reused helper, the generated test still fails after applying the golden patch.

\begin{minipage}{\linewidth}
\begin{lstlisting}[
    language=Java,
    caption={Generated reproduction test for Jackson Databind PR \#1923.},
    label={lst:jackson-1923-test}
]
public void testInterfaceType1899() throws Exception
{
    _testIllegalType(BogusInterface.class);
}
\end{lstlisting}
\end{minipage}

\begin{minipage}{\linewidth}
\begin{lstlisting}[
    language=Java,
    caption={Existing helper reused by the generated test.},
    label={lst:jackson-1923-helper}
]
protected void _verifySecurityException(Throwable t, String clsName)
    throws Exception
{
    _verifyException(t, JsonMappingException.class,
        "Illegal type",
        "to deserialize",
        "prevented for security reasons");
    verifyException(t, clsName);
}
\end{lstlisting}
\end{minipage}

\subsubsection{Effects on Existing Tests}
\label{app:failure-existing-tests}

\paragraph{Catch2 \#2719.}
\label{app:failure-catch2-2719}

\href{https://github.com/catchorg/Catch2/issues/2719}{\texttt{Catch2 \#2719}}
reports that an unexpected exception following \texttt{CHECKED\_ELSE} is not correctly reported as a test failure.
The agent generated four tests to reproduce this behavior.
All four tests failed before the golden patch and passed after it, successfully capturing the behavioral change introduced by the fix.
Nevertheless, the overall generated test set was classified as Non-success.

Listing~\ref{lst:catch2-2719-test} shows a representative generated test.
Although the generated tests themselves were successful after applying the golden patch, their addition changed the aggregate test output maintained by Catch2, as shown in Listing~\ref{lst:catch2-2719-output}.
Catch2 includes an existing \texttt{ApprovalTests} test that compares this output against a saved expected output.
The additional exception reports and changed summary counts therefore caused \texttt{ApprovalTests} to fail even after the golden patch was applied.

\begin{minipage}{\linewidth}
\begin{lstlisting}[
    language=C++,
    caption={Generated reproduction test for Catch2 \#2719.},
    label={lst:catch2-2719-test}
]
TEST_CASE( "Unexpected exception after CHECKED_ELSE(false) fails the test",
           "[.][failing][!throws]" ) {
    CHECKED_ELSE( false ) {}
    throw std::runtime_error( "unexpected exception" );
}
\end{lstlisting}
\end{minipage}

\begin{minipage}{\linewidth}
\begin{lstlisting}[
    language=diff,
    caption={Change in Catch2's aggregate test output after adding the generated tests.},
    label={lst:catch2-2719-output}
]
+ Unexpected exception after CHECKED_ELSE(false) fails the test
+ ...
+ due to unexpected exception with message:
+   unexpected exception

- test cases:  409 | 322 passed | 69 failed | 7 skipped | 11 failed as expected
+ test cases:  411 | 322 passed | 71 failed | 7 skipped | 11 failed as expected

- assertions: 2208 | 2048 passed | 128 failed | 32 failed as expected
+ assertions: 2211 | 2049 passed | 130 failed | 32 failed as expected
\end{lstlisting}
\end{minipage}

\subsection{Cross-Language Challenges}

\subsubsection{Implicit Setup Requirements}
\label{app:failure-implicit-setup}

\paragraph{Day.js \#1022.}
\label{app:failure-dayjs-1022}

\href{https://github.com/iamkun/dayjs/issues/1022}{\texttt{Day.js \#1022}}
reports that combining the RelativeTime and BadMutable plugins causes \texttt{fromNow()} to return an incorrect relative-time string.
The issue provides a concrete timestamp and the expected output \texttt{4 months ago}.
Reproducing this exact behavior in a test, however, requires not only using the reported timestamp but also initializing both plugins and controlling the current time used by \texttt{fromNow()}.

Listing~\ref{lst:dayjs-1022-success} shows a successful reproduction test generated by MSWE-agent.
The test initializes both plugins and fixes the current time using \texttt{MockDate}, allowing the reported input and expected output to be reproduced under the required conditions.
This test failed before the golden patch with \texttt{Infinity years ago} instead of \texttt{4 months ago}, and passed after the patch.

Other runs of the same agent did not complete an executable reproduction test despite receiving the same issue description.
For example, one run created a test file but left it empty, causing Jest to reject the suite both before and after the golden patch (see Listing~\ref{lst:dayjs-1022-incomplete}).
Another run similarly submitted an empty test file, while a further run submitted no new test.
Thus, the unsuccessful runs did not reach an executable test that instantiated the conditions required to reproduce the reported behavior.

\begin{minipage}{\linewidth}
\begin{lstlisting}[
    caption={Successful reproduction test for Day.js \#1022.},
    label={lst:dayjs-1022-success}
]
import MockDate from 'mockdate'
import dayjs from '../../src'
import relativeTime from '../../src/plugin/relativeTime'
import badMutable from '../../src/plugin/badMutable'

dayjs.extend(relativeTime)
dayjs.extend(badMutable)

beforeEach(() => {
  MockDate.set(new Date('2020-08-22T00:00:00.000Z'))
})

afterEach(() => {
  MockDate.reset()
})

it('calculates relative time correctly with BadMutable', () => {
  const time = dayjs(1588262400000).fromNow()

  expect(time).toBe('4 months ago')
})
\end{lstlisting}
\end{minipage}

\begin{minipage}{\linewidth}
\begin{lstlisting}[
    language=diff,
    caption={Incomplete reproduction test in an unsuccessful run for Day.js \#1022.},
    label={lst:dayjs-1022-incomplete}
]
Final submitted patch:
--- /dev/null
+++ b/test/plugin/relativeTimeWithBadMutable.test.js
@@ -0,0 +1 @@
+

Evaluation before and after the golden patch:
FAIL test/plugin/relativeTimeWithBadMutable.test.js
  Test suite failed to run

  Your test suite must contain at least one test.

Test Suites: 1 failed, 1 total
Tests:       0 total
\end{lstlisting}
\end{minipage}

\subsubsection{Loss of Intended Behavior or Executability after Revisions}
\label{app:failure-revisions}

\paragraph{Express \#3695.}
\label{app:failure-express-3695}

\href{https://github.com/expressjs/express/issues/3696}{\texttt{Express \#3696}}
requests that errors thrown by asynchronous request handlers be automatically forwarded to Express's error-handling middleware.
The corresponding benchmark instance is derived from
\href{https://github.com/expressjs/express/pull/3695}{\texttt{PR \#3695}}.
The agent initially generated a test using an \texttt{async} handler that throws an error, which correctly reproduces the reported behavior.

However, the initial test did not satisfy the repository's lint configuration.
As shown in Listing~\ref{lst:express-3695-revision}, the agent revised the handler from an \texttt{async function} that throws an error to an ordinary function that returns a rejected Promise.
Although this change resolved the lint error, it also changed the behavior exercised by the test.
Listing~\ref{lst:express-3695-gold} shows that the golden patch handles rejected Promises only when the request handler is an \texttt{AsyncFunction}.
The revised handler therefore bypasses the newly added behavior and continues to fail even after the golden patch is applied.
A controlled rerun confirmed that the initial test fails before the golden patch and passes afterward, whereas the revised test fails on both versions.

\begin{minipage}{\linewidth}
\begin{lstlisting}[
    language=diff,
    caption={Revision of the generated reproduction test for Express PR \#3695.},
    label={lst:express-3695-revision}
]
- router.get('/foo', async function(req, res, next){
-   throw new Error('promise error');
+ router.get('/foo', function(req, res, next){
+   return Promise.reject(new Error('promise error'));
  });
\end{lstlisting}
\end{minipage}

\begin{minipage}{\linewidth}
\begin{lstlisting}[
    language=diff,
    caption={Relevant behavior introduced by the golden patch for Express PR \#3695.},
    label={lst:express-3695-gold}
]
+ if (fn.constructor.name === 'AsyncFunction') {
+   return fn(req, res, next).catch(next)
+ }
\end{lstlisting}
\end{minipage}

\paragraph{Ponyc \#1051.}
\label{app:failure-ponyc-1051}

\href{https://github.com/ponylang/ponyc/issues/1050}{\texttt{Ponyc \#1050}}
reports that a parenthesized \texttt{return} can bypass a compiler diagnostic that should reject a terminal return at the end of a method.
The corresponding benchmark instance is derived from
\href{https://github.com/ponylang/ponyc/pull/1051}{\texttt{PR \#1051}}.
The agent initially generated a test that expects the compiler to report the corresponding diagnostic, as shown in Listing~\ref{lst:ponyc-1051-test}.
A controlled rerun confirmed that this test fails before the golden patch and passes afterward.

However, a later edit introduced a duplicate definition of an existing GoogleTest test.
As shown in Listing~\ref{lst:ponyc-1051-duplicate}, the final test file contains two definitions with the same test name.
The resulting C++ compilation error occurs before the generated reproduction test can be executed.
Consequently, the final submitted test suite fails to compile both before and after applying the golden patch, even though the initially generated reproduction test itself captured the intended behavioral change.

\begin{minipage}{\linewidth}
\begin{lstlisting}[
    language=C++,
    caption={Generated reproduction test for Ponyc PR \#1051.},
    label={lst:ponyc-1051-test}
]
TEST_F(BadPonyTest, ParenthesizedReturnAtEndOfMethod)
{
  const char* src =
    "actor Main\n"
    "  new create(env: Env) =>\n"
    "    (return)";

  TEST_ERRORS_1(src,
    "use return only to exit early from a method, not at the end");
}
\end{lstlisting}
\end{minipage}

\enlargethispage{2\baselineskip}
\begin{minipage}{\linewidth}
\begin{lstlisting}[
    caption={Duplicate test definition introduced by a later edit in Ponyc PR \#1051.},
    label={lst:ponyc-1051-duplicate}
]
TEST_F(BadPonyTest, TypeAliasRecursionThroughTypeParameterInTuple)
{
    ...
}

TEST_F(BadPonyTest, TypeAliasRecursionThroughTypeParameterInTuple)
{
    ...
}

test/libponyc/badpony.cc:212:8: error:
redefinition of
'class BadPonyTest_TypeAliasRecursionThroughTypeParameterInTuple_Test'

test/libponyc/badpony.cc:200:8: note:
previous definition of
'class BadPonyTest_TypeAliasRecursionThroughTypeParameterInTuple_Test'
\end{lstlisting}
\end{minipage}

% TODO: Add supplementary material.

\end{document}